\documentclass{aa}

\usepackage{graphicx}
\usepackage{subfig}

\usepackage{txfonts}
\usepackage[]{hyperref}
\usepackage{url,natbib}

\bibpunct{(}{)}{;}{a}{}{,}    

\usepackage{caption,booktabs,array}

\hypersetup{
  colorlinks=true,   
  urlcolor=blue,     
  linkcolor=blue,     
  citecolor = blue
}
\usepackage{siunitx}
\usepackage{comment}
\usepackage[]{xcolor} 
\usepackage{mathabx}

\usepackage{float}

\newcommand{\ms}{\mbox{m\,s$^{-1}$}}
\newcommand{\kms}{\mbox{km\,s$^{-1}$}}
\newcommand{\logg}{\mbox{$\log g$}}
\newcommand{\teff}{\mbox{$T_{\rm eff}$}}
\newcommand{\MEarth}{\mbox{$\mathrm{M}_\Earth$}}

\newif\ifshowcomOD
\showcomODtrue

\newif\ifshowcomTS
\showcomTStrue

\usepackage[normalem]{ulem}

\begin{document} 

   \title{The HD\,137496 system: \\ A dense, hot super-Mercury and a cold Jupiter\thanks{Tables \ref{table:measur_rvs} and \ref{table:photo} are only available in electronic form at the CDS via anonymous ftp to \texttt{cdsarc.u-strasbg.fr (130.79.128.5)} or via \url{http://cdsweb.u-strasbg.fr/cgi-bin/qcat?J/A+A/}.}}

   \author{T. A. Silva\inst{\ref{IA-Porto}, \ref{UPorto}\thanks{\email{Tomas.Silva@astro.up.pt}}}
           \and O.D.S Demangeon\inst{\ref{IA-Porto},\thanks{\email{olivier.demangeon@astro.up.pt}}}
           \and S. C. C. Barros \inst{\ref{IA-Porto}, \ref{UPorto}}
           \and D. J. Armstrong \inst{\ref{Warwick}, \ref{CentreWarwick}}
           \and J. F. Otegi \inst{\ref{Geneva}, \ref{Zurich}}
           \and D. Bossini \inst{\ref{IA-Porto}}
           \and E. Delgado Mena \inst{\ref{IA-Porto}}
           \and S. G. Sousa \inst{\ref{IA-Porto}}
           \and V. Adibekyan \inst{\ref{IA-Porto}, \ref{UPorto}}
           \and L. D. Nielsen \inst{\ref{Geneva}, \ref{Oxford}}
           \and C. Dorn \inst{\ref{Zurich}}
           \and J. Lillo-Box \inst{\ref{INTA}}
           \and N. C. Santos \inst{\ref{IA-Porto}, \ref{UPorto}}
           \and S. Hoyer \inst{\ref{Marseille}}
           \and K. G.\ Stassun \inst{\ref{USA}}
           \and J. M. Almenara \inst{\ref{Grenoble}}
           \and D. Bayliss \inst{\ref{Warwick}, \ref{CentreWarwick}}
           \and D. Barrado \inst{\ref{INTA}}
           \and I. Boisse \inst{\ref{Marseille}}
           \and D. J. A. Brown \inst{\ref{Warwick}, \ref{CentreWarwick}}
           \and R. F. D\'iaz \inst{\ref{Argentina}}
           \and X. Dumusque \inst{\ref{Geneva}}
           \and P. Figueira \inst{\ref{ESO}, \ref{IA-Porto}}
           \and A. Hadjigeorghiou \inst{\ref{Warwick}, \ref{CentreWarwick}}
           \and S. Hojjatpanah \inst{\ref{Marseille}}
           \and O. Mousis \inst{\ref{Marseille}}
           \and A. Osborn \inst{\ref{Warwick}, \ref{CentreWarwick}}
           \and A. Santerne \inst{\ref{Marseille}}
           \and P. A. Str{\o}m \inst{\ref{Warwick}}
           \and S. Udry \inst{\ref{Geneva}}
           \and P. J. Wheatley \inst{\ref{Warwick}, \ref{CentreWarwick}}
           }

   \institute{Instituto de Astrof\'isica e Ci\^{e}ncias do Espa\c co, Universidade do Porto, CAUP, Rua das Estrelas, 4150-762 Porto, Portugal
\label{IA-Porto}
             \and
             Departamento de Física e Astronomia, Faculdade de Ciências, Universidade do Porto, Rua do Campo Alegre, 4169-007 Porto, Portugal. \label{UPorto}
             \and
             Department of Physics, University of Warwick, Gibbet Hill Road, Coventry CV4 7AL, UK. \label{Warwick}
             \and
             Centre for Exoplanets and Habitability, University of Warwick, Gibbet Hill Road, Coventry CV4 7AL, UK. \label{CentreWarwick}
             \and
             Geneva Observatory, University of Geneva, Chemin Pegasi 51, Versoix CH-1290, Switzerland. \label{Geneva}
             \and 
             Institute for Computational Science, University of Zurich,Winterthurerstr. 190, CH-8057 Zurich, Switzerland. \label{Zurich}
             \and
             Department of Physics, University of Oxford, OX13RH, Oxford, UK. \label{Oxford}
             \and 
             Centro de Astrobiolog\'ia (CAB, CSIC-INTA), Depto. de Astrof\'isica, ESAC campus 28692 Villanueva de la Ca\~nada (Madrid), Spain. \label{INTA}
             \and 
             Aix Marseille Univ, CNRS, CNES, LAM, Marseille, France. \label{Marseille}
             \and 
             Department of Physics and Astronomy, Vanderbilt University, Nashville, TN 37235, USA. \label{USA}
             \and 
             Univ. Grenoble Alpes, CNRS, IPAG, F-38000 Grenoble, France. \label{Grenoble}
             \and
             International Center for Advanced Studies (ICAS) and ICIFI (CONICET), ECyT-UNSAM, Campus Miguelete, 25 de Mayo y Francia, (1650) Buenos Aires, Argentina. \label{Argentina}
             \and 
             European Southern Observatory, Alonso de Córdova 3107, Santiago, Chile. \label{ESO}
             }

   \date{Received September 15, 1996; accepted March 16, 1997}

 
  \abstract
   {Most of the currently known planets are small worlds with radii between that of the Earth and that of Neptune. The characterization of planets in this regime shows a large diversity in compositions and system architectures, with distributions hinting at a multitude of formation and evolution scenarios. However, many planetary populations, such as high-density planets, are significantly under-sampled, limiting our understanding of planet formation and evolution. }
  {NCORES is a large observing program conducted on the HARPS high-resolution spectrograph that aims to confirm the planetary status and to measure the masses of small transiting planetary candidates detected by transit photometry surveys in order to constrain their internal composition.
  }
  {Using photometry from the K2 satellite and radial velocities measured with the HARPS and CORALIE spectrographs, we searched for planets around the bright (V$_\textrm{mag}$ = 10) and slightly evolved Sun-like star HD\,137496.
  }
   {We precisely estimated the stellar parameters, $M_*$ =  1.035 $\pm$ 0.022 $M_\odot$, $R_*$ =  1.587 $\pm$ 0.028 $R_\odot$, $T_\text{eff}$ =  5799 $\pm$ 61 K,  together with the chemical composition (e.g. [Fe/H] = -0.027 $\pm$ 0.040 dex) of the slightly evolved star. We detect two planets orbiting HD\,137496.
   The inner planet, HD\,137496\,b, is a super-Mercury (an Earth-sized planet with the density of Mercury) with a mass of
   $M_b$ = 4.04  $\pm$  0.55 $M_\oplus$, a radius of $R_b = 1.31_{-0.05}^{+0.06}\,R_\oplus,$ and a density of $\rho_b = 10.49_{-1.82}^{+2.08}$ $\mathrm{g \, cm^{-3}}$. With an interior modeling analysis, we find that the planet is composed mainly of iron, with the core representing over 70\% of the planet's mass ($M_{core}/M_{total} = 0.73^{+0.11}_{-0.12}$). 
   The outer planet, HD\,137496\,c, is an eccentric ($e$ = 0.477  $\pm$  0.004), long period ($P$ = $479.9_{-1.1}^{+1.0}$ days) giant planet ($M_c\sin i_c$ = 7.66\,$\pm$\,0.11\,$M_{Jup}$) for which we do not detect a transit.}
   {HD\,137496\,b is one of the few super-Mercuries detected to date. The accurate characterization reported here enhances its role as a key target to better understand the formation and evolution of planetary systems. The detection of an eccentric long period giant companion also reinforces the link between the presence of small transiting inner planets and long period gas giants.
}


   \maketitle
%

\section{Introduction}\label{sect:intro}

In our quest to better understand planet frequency and formation, few missions have been as influential as \textit{Kepler} \citep{2010Sci...327..977B}, with upwards of 2500 validated planets\footnote{See the \href{http://exoplanet.eu/catalog/}{Extrasolar Planets Encyclopaedia} \citep{2011A&A...532A..79S} or the \href{https://exoplanetarchive.ipac.caltech.edu/cgi-bin/TblView/nph-tblView?app=ExoTbls&config=cumulative&constraint=koi_pdisposition+like+\%27CANDIDATE\%27}{NASA Exoplanet Archive} \citep{akeson2013}.} and many others awaiting confirmation. 
More than ten years have passed since its launch, and astronomers are still able to reap the benefits of its data. In 2013, a second failure occurred on \textit{Kepler}'s reaction wheels. Despite this technical problem, \textit{Kepler} continued operations but surveyed the ecliptic plane for periods of around 80 days called campaigns. This new operational setup of the \textit{Kepler} satellite became known as the \textit{K2} mission \citep{2014PASP..126..398H}.

Together with other space-based missions and ground-based telescopes, \textit{Kepler} and \textit{K2} have discovered a large number of planets, many of them in configurations foreign to our Solar System. Not only is the architecture of the orbits distinct, the physical characteristics of these planets are also significantly different, with a broad spectrum of sizes, masses, and plausible compositions.  
Together with former results of Doppler radial velocity surveys \citep[e.g.,][]{2013A&A...549A.109B, 2014Natur.513..328M}, over the last decade it has become clear that super-Earths and sub-Neptunes, planets with a radius between that of Earth and and that of Neptune, are ubiquitous in our Galaxy despite not being found in our Solar System \citep{2012ApJS..201...15H, 2018ApJ...860..101Z}.

One of the most striking results for this population is the presence of a radius gap \citep{2017AJ....154..109F}, where observed planets appear to prefer a radius close to $\sim$1.3$R_\oplus$ or to $\sim$2.6$R_\oplus$. 
This gap hints at two separate populations of exoplanets, where the smaller radius planets are thought to be predominantly rocky, while planets with a radius above the gap are expected to have lower densities and extended atmospheres rich in volatiles. In order to explain this bimodal distribution (super-Earths and sub-Neptunes), multiple theoretical models have been suggested. One of the most prominent hypotheses is that this distribution is a result of photoevaporation \citep{2012ApJ...761...59L, 2017ApJ...847...29O}. Photoevaporation, the erosion of the planetary atmosphere by the stellar irradiation, can strip the atmosphere of a planet down to its core, while less irradiated and/or heavier planets are able to retain their atmospheres and sustain a bigger radius. A concurrent explanation is that the atmosphere of the planets can escape as a result of the thermal emission generated by the planetary core \citep{2019MNRAS.487...24G}. Recent works also propose that the stellar environment of a planetary system plays a crucial role in the observed radius distribution \citep{2020ApJ...905L..18K}. The exoplanet community is now trying to characterize more planets in this regime to identify which are the best hypotheses \citep[e.g.,][]{2018MNRAS.479.5303L, fulton2018california, van2018asteroseismic, cloutier2020evolution, 2020AJ....160...22C, 2021A&A...648A..75S, 2021A&A...649A..26L}. 

Within the rocky planet population, a variety of compositions are estimated, from seemingly Earth-like compositions \citep[e.g.,][]{dressing2015} to high-density Mercury-like planets \citep[e.g.,][]{2018NatAs...2..393S}. When looking at planets with interior compositions resembling that of Mercury, additional formation and evolution scenarios are expected to take place to give origin to an iron-enhanced planet. Such hypotheses include that some planets are initially formed in iron-rich environments \citep[e.g.,][]{2013ApJ...769...78W, 2017A&A...608A..94S, 2020ApJ...901...97A, 2021arXiv210212444A}, suffer from giant impacts \citep{2010ApJ...712L..73M, inamdar2015formation}, or even experience impact erosion from planetesimals \citep{2009M&PS...44.1095S, wyatt2020susceptibility}. Planetary dynamics can also play a significant role in the transition from gas giants to high-density planets; this includes tidal disruption effects and Roche-Lobe Overflow of matter and angular momentum transfers \citep{2003ApJ...588..509G,  2010Natur.463.1054L, 2013ApJ...773L..15R, 2014ApJ...793L...3V, 2015ApJ...813..101V, 2017ApJ...835..145J}. The detection of elusive Mercury-like exoplanets and the study of their system architectures can provide unique and important clues regarding the formation of these planets. 

The large number of exoplanets discovered with \textit{Kepler} and \textit{K2} shed new light on the processes shaping these planets while also enabling more complete studies on the architecture of planetary systems. 
A recent study using \textit{Kepler} data \citep{2018AJ....156...92Z} found that there is a positive correlation between the presence of hot super-Earths and "cold" Jupiters in the same system. 
This work was later corroborated by \cite{2019AJ....157...52B}, where the authors correspondingly find that $\sim$ 1/3 of close-in systems of super-Earths and sub-Neptunes possess exterior giants companions (0.5-20  $M_{Jup}$ at 1–20 AU). These results imply that the presence of outer gas giants does not prevent the formation of super-Earths. 
On the other hand, \cite{2020A&A...643A..66B} showed in their simulations that the eccentricity of these massive companions may play a role in the presence of super-Earths, for which systems with eccentric Jupiter-type planets are less likely to host inner super-Earths.
Identifying and studying such correlations is important to understanding the origin of these planets, as it allows astronomers to test different models of planet formation and migration and gather insights into the dynamics and the architectures of planetary systems \citep{2012ApJ...761...92F}.

In this paper, we present the discovery and characterization of such a system, a new multi-planetary system around HD\,137496, alias K2-364 (EPIC\footnote{EPIC stands for Ecliptic Plane Input Catalog} 249910734), a V$_\textrm{mag}$=10, mid-G star near the main sequence turn-off. HD\,137496\,b is a short period super-Mercury, and HD\,137496\,c is a Jupiter-sized planet. HD\,137496\,b was first identified based on photometric data from \textit{K2} Campaign 15 and confirmed with radial velocity (RV) observations from the HARPS and CORALIE instruments. These measurements revealed the presence of HD\,137496\,c, the second planet in the system. This system's radial velocities were obtained within the scope of the NCORES program  (ID 1102.C-0249, PI Armstrong). This HARPS program focuses on improving our understanding of the origin and internal composition of such close-in planets \citep[e.g.,][]{2020MNRAS.492.5399N, 2020Natur.583...39A, 2021MNRAS.502.4842O}.

This paper is structured as follows. In Section \ref{section:photometry}, we describe the transit discovery and characterization from the \textit{K2} photometry. We then check for photometric contamination in Section \ref{sec:cont}. Next, we discuss the RV analysis and stellar activity indicators in Section \ref{section:rvs}. The stellar properties are provided in Section \ref{section:stellar}.
The joint analysis of the RVs and photometry are reported in Section \ref{sec:joint}, and the discussion and conclusions are given in Section \ref{sec:conc}.

\section{\textit{K2} photometry} \label{section:photometry}

HD\,137496 was observed by the \textit{Kepler} spacecraft during Campaign 15 of the \textit{K2} mission from 23 August 2017 to 19 November 2017 - programs: GO15009 (PI: Charbonneau), GO15021 (PI: Howard) and GO15028 (PI: Cochran). This target is located at RA: 15:26:58.07, Dec: -16:30:31.68 (J2000) and was observed in the long ($\sim$ 30 min) cadence mode.

The raw data were extracted through the Mikulski Archive for Space Telescopes (MAST)\footnote{ \url{ http://archive.stsci.edu/k2/epic/search.php}.} portal. These include the target pixel files, the simple aperture photometry (SAP) flux, and the more processed version of SAP with artifact mitigation called the pre-search data conditioning (PDCSAP) flux. 

\subsection{Detection of a transit signal with a period of 1.62\,days}

As most \textit{K2} light curves, the SAP and PDSCAP light curves suffer from a high noise level with a 12 hr period. This is  due to the rolling motion of the satellite caused by solar pressure, which results from the loss of two reaction wheels and the periodic corrections of this motion. For this reason, we used the Planet candidates from OptimaL Aperture Reduction (POLAR) software \citep{2016A&A...594A.100B}, which was specifically developed to mitigate this noise source in order to assess the presence of transit signal in the light curve.
Using the box-fitting least-squares \citep[BLS; ][]{2002A&A...391..369K} algorithm implemented within the POLAR pipeline \citep[see][for more details]{2016A&A...594A.100B}, we detected the presence of a transit signal with a period of 1.62\,days in the light curve of HD\,137496.

\subsection{Comparison of POLAR, K2SFF, and EVEREST photometric performances}

\begin{figure}[!htb]
        \centering
        \includegraphics[width=\linewidth]{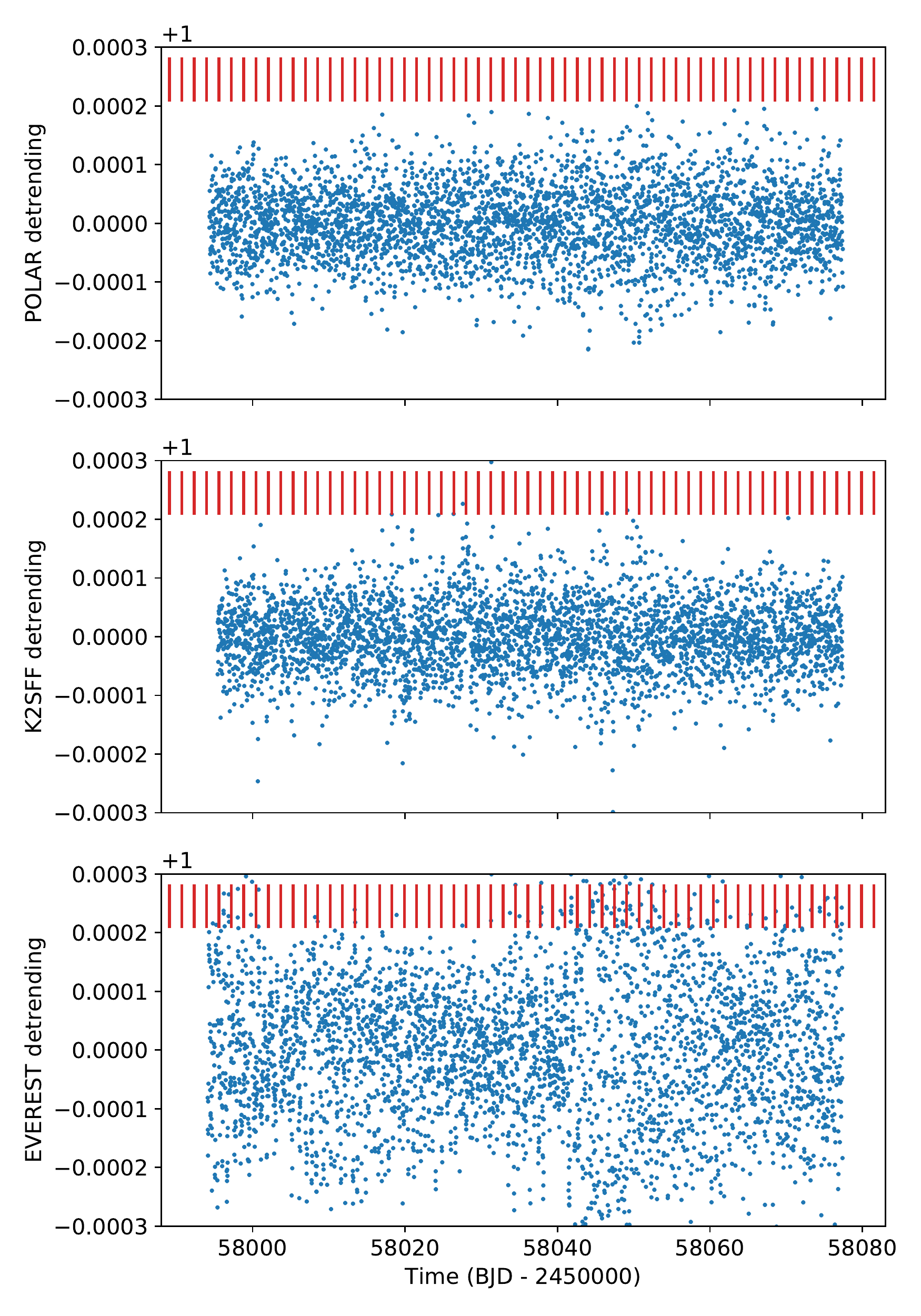}
        \caption{Normalized flux from K2 photometry with systematics corrected via the POLAR pipeline (top panel), K2SFF pipeline (middle panel), and EVEREST pipeline with a Savitzky–Golay filtering (bottom panel). The red lines indicate the expected transit positions for the short-period planet (HD\,137496 b).
}
    \label{fig:3_detrending_all_time}
\end{figure}

Similarly to POLAR, other software has been developed to provide additional noise mitigation for data provided by the \textit{K2} mission. Such software includes K2SFF \citep{2014PASP..126..948V}, which uses self-flat-fielding methods (as POLAR does), and EVEREST \citep{2016AJ....152..100L, 2018AJ....156...99L}, which uses pixel level decorrelation; these have made their light curves publicly available. In order to select the best light curve for the characterization of the discovered transit signal, we compared the photometric precision of the three light curves provided by POLAR\footnote{POLAR provides activity-corrected light curves as well as instrumental detrended light curves.}, K2SFF, and EVEREST and present this in Fig. \ref{fig:3_detrending_all_time}. 

\subsubsection{Pre-treatment of the light curves}

Both POLAR and K2SFF light curves are provided already detrended. This means that long-term variations from residual instrumental noises or stellar variability have been removed and the light curves are flat and ready for transit detection and analysis. This is not the case with the EVEREST light curves. Consequently, we detrended the EVEREST light curve using an adapted Savitzky–Golay filter \citep{1964AnaCh..36.1627S}. In our analysis, we used a second-degree polynomial and a 1.45 day window. 
Before applying the filter, we masked the transits identified using the period, the time of inferior conjunction, and the duration of the transit provided by the POLAR BLS search.
Contrary to other implementations of the Savitzky–Golay filter (for example, \texttt{scipy.signal.savgol\_filter} provided within the widely used Python package \texttt{scipy}), our implementation allows for unevenly sampled data\footnote{If the data points are within half the Savtzky–Golay filter window width from the beginning or end of the observation sequence, the coefficients of the closest valid point were used instead of the usual full window width.}. This is critical to properly filtering the regions containing masked transit signals. We detrended the original EVEREST light curve, dividing it by the result of the Savitzky–Golay filtering.

Finally, for all light curves, the data points not flagged as \texttt{QUALITY}=0\footnote{More details on the quality assessment can be found in \cite{2018AJ....156...99L}.} in the raw data were removed.
During the first days of observation, there is also an abrupt excess on the flux. This effect is observed in the photometry of different targets of \textit{K2} Campaign 15 and can be explained by a thermal anomaly on \textit{Kepler}'s spacecraft \citep{2020A&A...636A..89H}. For this reason, we removed all observations taken before $t$ = 2\,457\,994.328 BJD\footnote{Barycentric Julian dates throughout this work are computed
from the coordinated universal time (UTC) in the case of RVs and from the barycentric dynamical time (TDB) when regarding the Kepler light curve.}. 

\subsubsection{Photometric precision}

In order to measure the photometric precision of each light curve, we computed the 2\,hr\footnote{Given the 2.17 hr duration of the detected transit, the 2 hr CDPP is perfectly suited to select the best light curve for our analysis.}
CDPP \citep[combined differential photometric precision; ][]{2012PASP..124.1279C} using the implementation provided with the EVEREST pipeline. The results are a CDPP of 20.78 ppm for the POLAR light-curve, 19.89 ppm for the K2SFF one and 28.47 ppm for the EVEREST detrended light curve.

\begin{figure}[!htb]
        \centering
        \includegraphics[width=\linewidth]{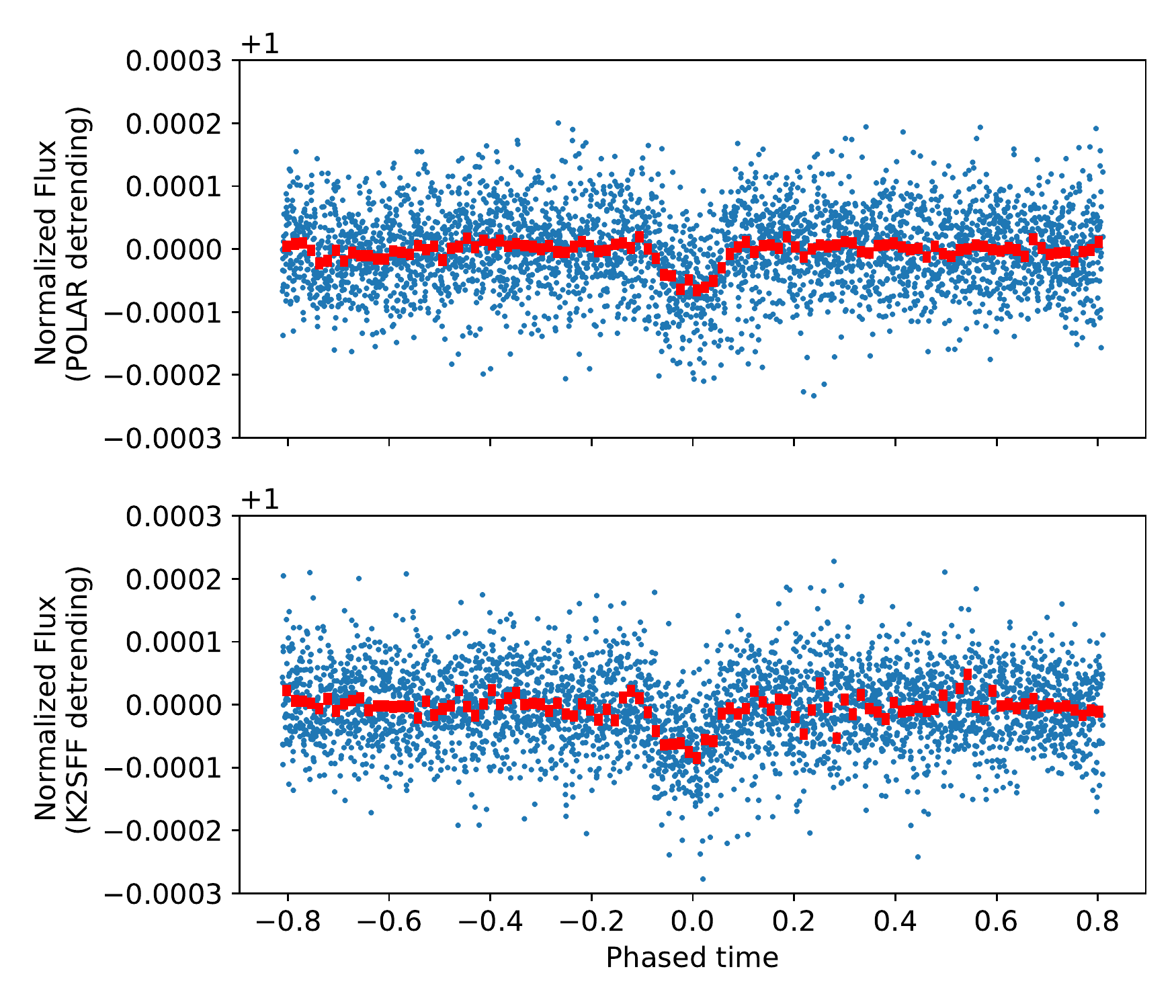}
        \caption{Phase-folded POLAR (top) and K2SFF (bottom) photometric transit data (blue points). Represented in red are the 30 min bins.
}
    \label{fig:plot_3subs_phase_curves}
\end{figure}

The POLAR and K2SFF methods have similar 2h-CDPP values and, from a visual inspection, similar phase-folded light curves (see Fig. \ref{fig:plot_3subs_phase_curves}). With both detrending methods presenting equivalent results, we decided to use the POLAR\footnote{POLAR light curves are provided upon request to the authors.} detrended light curve (Table \ref{table:photo}) for the rest of our analysis.

\section{Photometric contamination} \label{sec:cont}

From the GAIA DR3 \citep{2016A&A...595A...1G, 2021A&A...649A...1G} catalog, we find that two additional stars\footnote{GAIA DR3 ID sources: 6258809897751679360 and 6258809794673492864} fall within the K2 aperture. Their PSFs may contaminate the pixels used by POLAR to extract the photometry. However, the combined flux of these stars is four orders of magnitude smaller than the flux of our target and the impact on the planetary parameter is thus negligible.

\begin{figure}[!htb]
        \centering
        \includegraphics[width=\linewidth]{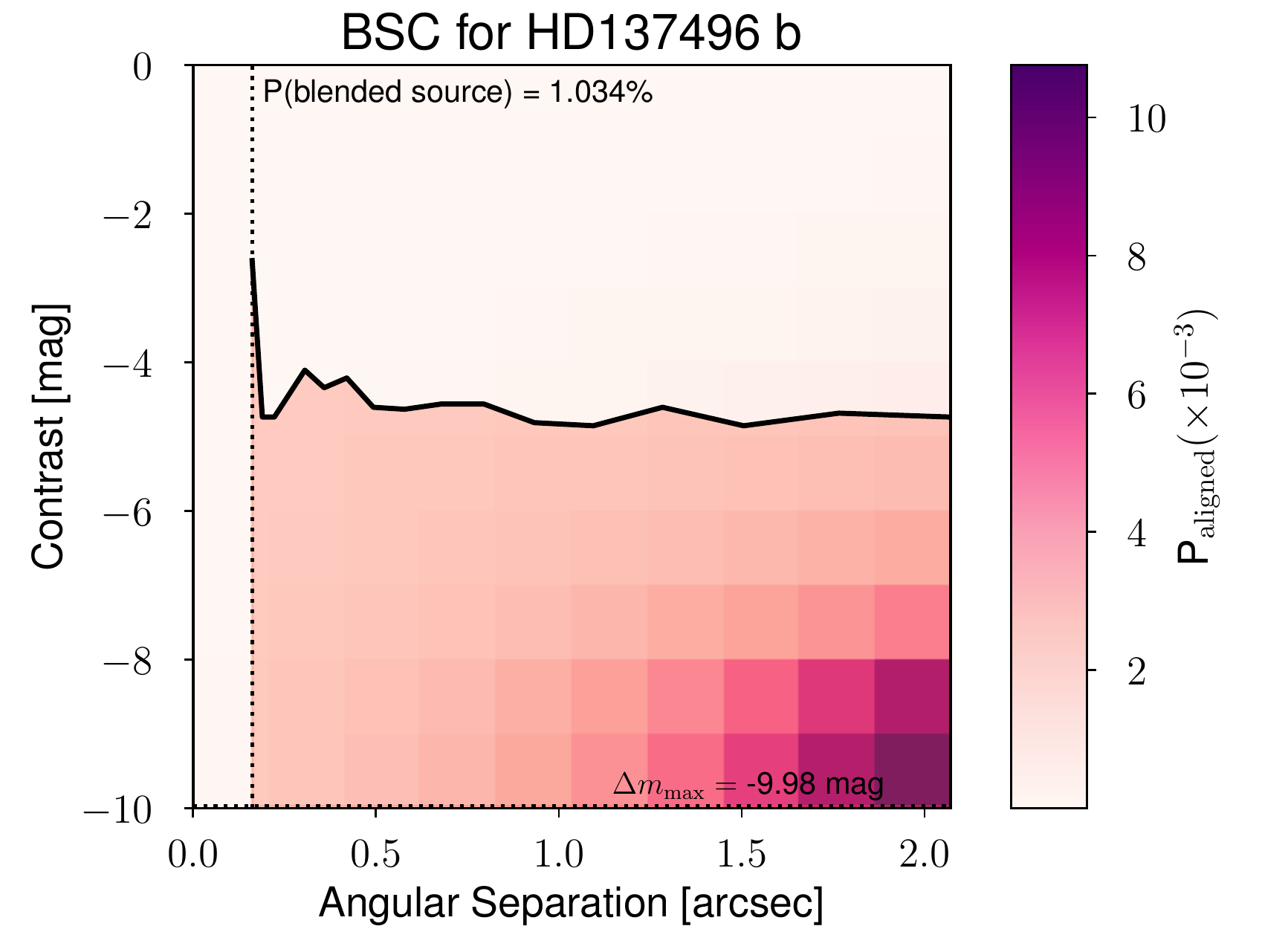}
        \caption{Probability of a chance-aligned source at the target location (more information in Section \ref{sec:cont}).}
    \label{fig:astralux}
\end{figure}

We observed HD\,137496 with the AstraLux lucky-imaging instrument \citep{hormuth08} at Calar Alto Observatory (Almer\'ia, Spain) under Prog. ID F20-2.2-014 (PI: J. Lillo-Box). The observations were carried out on the night of 7 July 2020 under good observing conditions with seeing around 1 arcsec. By using the lucky-imaging technique, we obtained 42730 frames with a 10\,ms exposure time using the Sloan Digital Sky Survey z filter (SDSSz) and windowing the total field of view down to a 6$\times$6 arcsec frame. This datacube of images is then reduced by the instrument pipeline (described in \citealt{hormuth08}) by performing the basic bias subtraction and flat-fielding and then selecting the top 10\% of frames according to the Strehl ratio metrics \citep{strehl1902}. Then, these best-quality images are aligned and combined together to form a final high-spatial resolution image. No additional sources are found in the final image within the sensitivity limits calculated with our own developed \texttt{astrasens} package\footnote{\url{https://github.com/jlillo/astrasens}}. The procedure is fully described in \cite{lillo-box12,lillo-box14b}. The final sensitivity curve is shown in Fig.~\ref{fig:astralux}. As shown, we can reach contrast magnitudes of 5 mag at 0.2 arcsec from the main target. 

From our high-resolution image, we can also estimate the probability of an undetected blended companion that could mimic the transit depth detected. We call this the blended source confidence (BSC), and the steps for estimating this probability are fully described in
\cite{lillo-box14b}. We used a Python implementation of this technique (\texttt{bsc}, by J. Lillo-Box), which uses the TRILEGAL5 galactic model (v1.6 \citealt{girardi12}) to retrieve a simulated source
population of the region around the corresponding target\footnote{This is done by using the Python implementation of \texttt{astrobase} by \citealt{astrobase}.}. From this simulated population, we derive the density of stars around the target position (radius $r=1^{\circ}$) with the associated probability of chance-alignment at a given contrast magnitude and separation. We used the default parameters for the bulge, halo, thin and thick disks, and the lognormal initial mass function from \cite{chabrier01}. The contrast curve was then used to constrain this parameter space and estimate the final probability of an undetected blended companion. We considered potentially contaminant sources as those with a maximum contrast magnitude of $\Delta m_{\rm max}=-2.5\log{\delta}$, with $\delta$ defined as the transit depth of the candidate planet in the Kepler band. In this case, the transit depth of 6.8 ppt corresponds to a maximum contrast magnitude for relevant contaminants of $\Delta m_{\rm max} = 5$~mag. This represents the maximum magnitude that a blended star can have to mimic this transit depth. We converted the depth in the Kepler passband to the AstraLux SDSSz filter by simple conversions using the KIC catalog and found that $\Delta m_{\rm Kep} = 0.920\Delta m_{\rm SDSSz}$. The result of this calculation provides a BSC of 1.034\%.

\section{Radial velocities} \label{section:rvs}

\subsection{HARPS}
Radial velocity measurements of HD\,137496 were obtained with the HARPS spectrograph (R=115,000) mounted on the 3.6m telescope at ESO's La Silla Observatory \citep{2003Msngr.114...20M}. A total of 142 observations were taken between 1 March 2019 and 1 March 2021 in the high-accuracy mode, as part of the NCORES large program (ID 1102.C-0249, PI Armstrong). An exposure time of 1500s was used, giving a signal-to-noise ratio (S/N) of $\sim$70 per pixel. Typically, the star was observed two-to-three times per night. The data were reduced with the standard HARPS pipeline. RV measurements were derived using a weighted cross-correlation function (CCF) method using a G2V  template \citep{1996A&AS..119..373B,2002Msngr.110....9P}. The bisector inverse slope (BIS) and the full width at half maximum (FWHM) were measured using the methods of \cite{2011A&A...528A...4B}.

\subsection{CORALIE}
HD\,137496 was also observed with the CORALIE spectrograph (R=60,000) mounted on the 1.2m Swiss Euler telescope at La Silla Observatory \citep{CORALIE}. Thirty observations were taken between 9 July 2019 and 21 August 2021.  An exposure time of 1200s was used, yielding a S/N of $\sim$25 per pixel (at 5500\AA). Similarly to HARPS, the RV measurements were obtained through cross-correlation of the spectra with a binary G2V template \citep{1996A&AS..119..373B,2002Msngr.110....9P}. 
The FWHM, BIS, and other line-profile diagnostics were computed using the standard CORALIE DRS. 
Both CORALIE and HARPS spectra and derived RVs will be made public through the DACE portal\footnote{\url{https://dace.unige.ch/radialVelocities/?pattern=HD\%20137496}} hosted at the University of Geneva \citep{2015ASPC..495....7B}.

\subsection{RV analysis}

\begin{figure}[!htb]
        \centering
        \includegraphics[width=\linewidth]{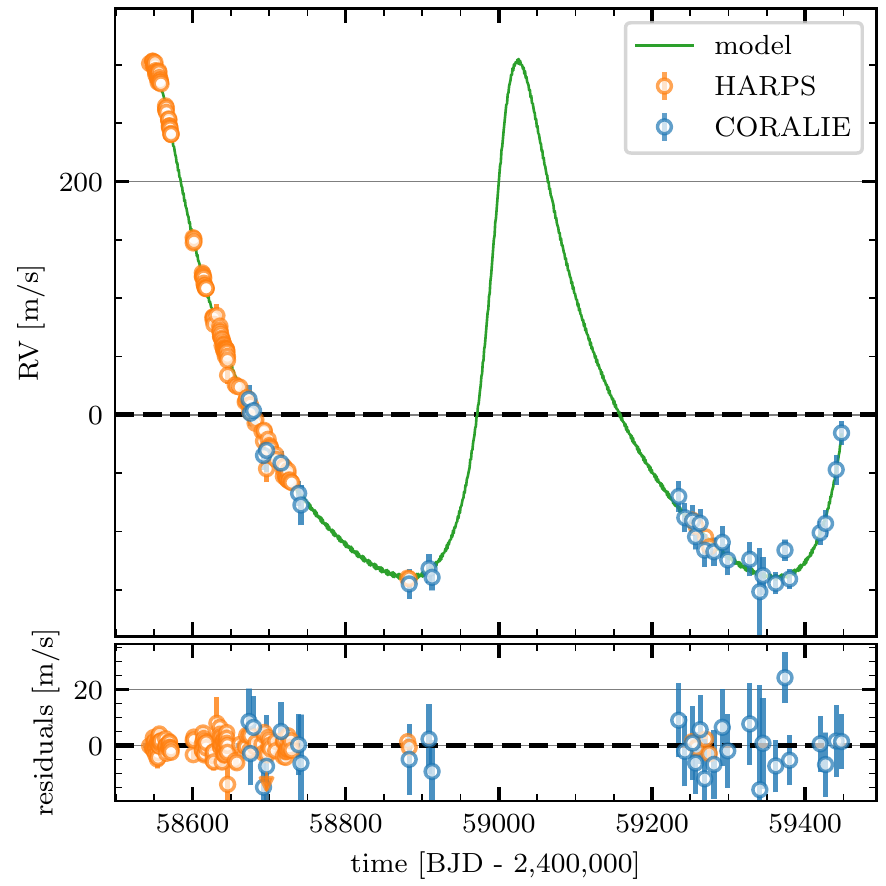}
        \caption{HARPS (orange) and CORALIE (blue) radial velocities. The calculation of the fit model (green) is explained on Section \ref{sec:joint}. 
        } 
    \label{fig:plot_rvs_harps_coralie}
\end{figure}

In Figure \ref{fig:plot_rvs_harps_coralie}, we present our RV time series. As is clearly seen, the data show a long-term and high-amplitude trend (semi-amplitude of $\sim200\,\ms$), typical of the signature of a long period giant planet.

{\renewcommand{\arraystretch}{1.2}

\begin{table*}
\caption{RVs and stellar activity indicators measured with HARPS and CORALIE.}
\label{table:measur_rvs}
\centering
\begin{tabular}{ccccccc}
\hline\hline
\text{BJD - 2400000} & RV (\ms) & FWHM (\ms)  & BIS (\ms) & ... & Instrument \\ \hline
58543.91 & $-2687.67 \pm 1.28$ & $7246.28 \pm 10.25$  & $11.39 \pm 1.8$ &  ... & HARPS \\
58547.80 & $-2685.88 \pm 1.44$ & $7247.69 \pm 10.25$  & $7.41 \pm 2.0$ &  ... & HARPS \\
58547.90 & $-2685.67 \pm 1.58$ & $7242.47 \pm 10.24$  & $8.42 \pm 2.2$ &  ... & HARPS \\
... & ... & ... & ... & ...  & ... \\

\hline
\\\end{tabular}

\tablefoot{The complete table will be made available in machine-readable form in the online journal.
}

\end{table*}}

\begin{figure*}[!htb]
        \centering
        \includegraphics[width=\linewidth]{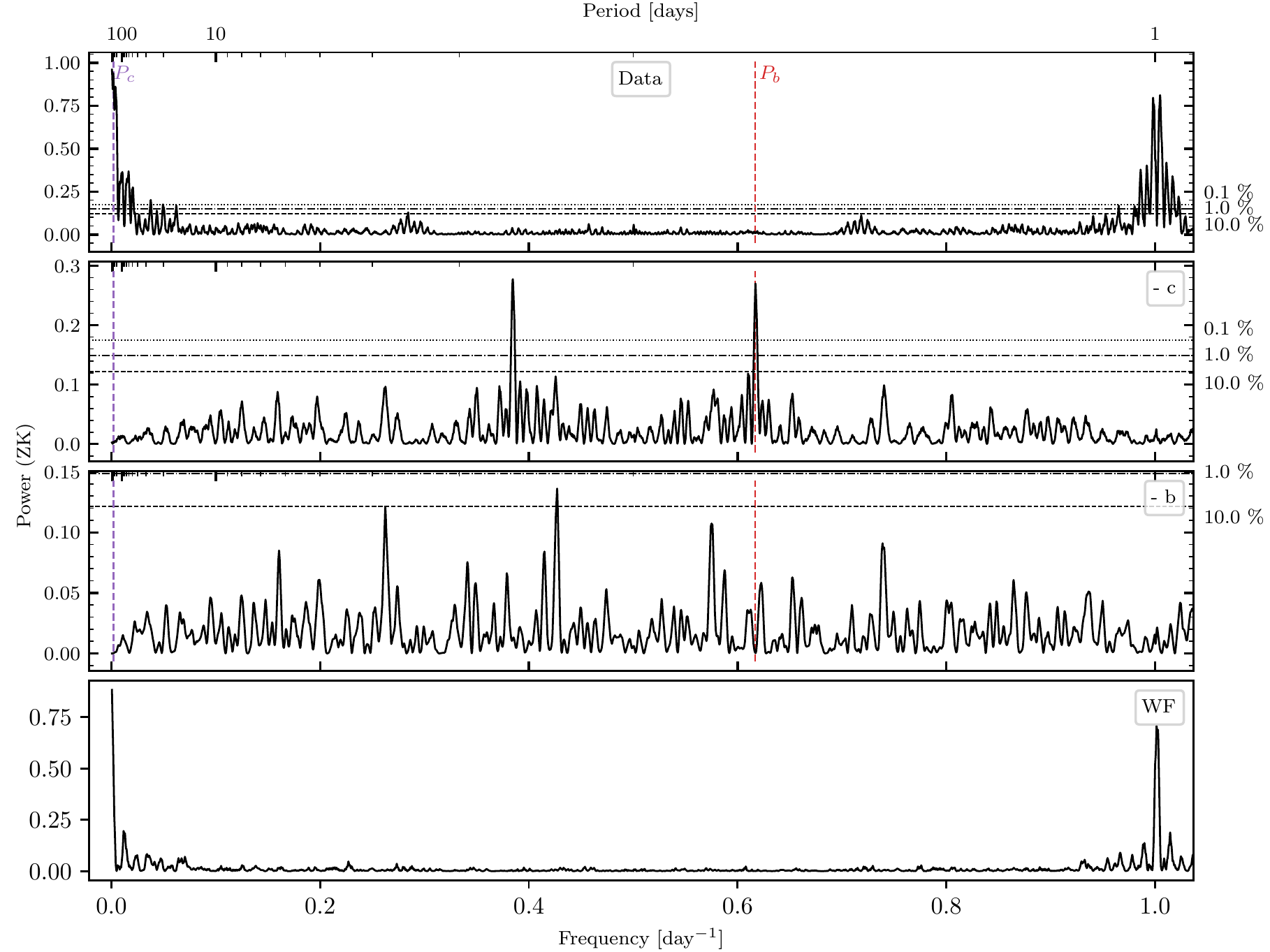}
        \caption{Generalized Lomb-Scargle periodogram of the RVs before (top) and after the fitting and removal of the signal from planet c (second from the top) and planet b (third from the top). The levels corresponding to a FAP of 10\%, 1\%, and 0.1\% are represented by the horizontal dashed lines. In the bottom panel, we show the GLSP of the window function, W($\nu$).
}
    \label{fig:GLS_planet_plot}
\end{figure*}

In the top panel of Fig. \ref{fig:GLS_planet_plot}, we plot the generalized Lomb-Scargle periodogram \citep[GLSP,][]{1976Ap&SS..39..447L, 1982ApJ...263..835S} of the RV time series. We notice that a high-significance, low false-alarm probability (FAP) peak is observed around a period of 500 days. After fitting a Keplerian to this signal and removing it from the RVs, the GLSP shown in the second panel of Fig. \ref{fig:GLS_planet_plot} displays a significant peak at the period detected in \textit{K2} photometric data by the BLS, $P=1.62$ days. The other significant peak (period of 2.6 days) in this GLSP corresponds to an alias of this 1.62 days peak. 

\begin{figure}[!htb]
        \centering
        \includegraphics[]{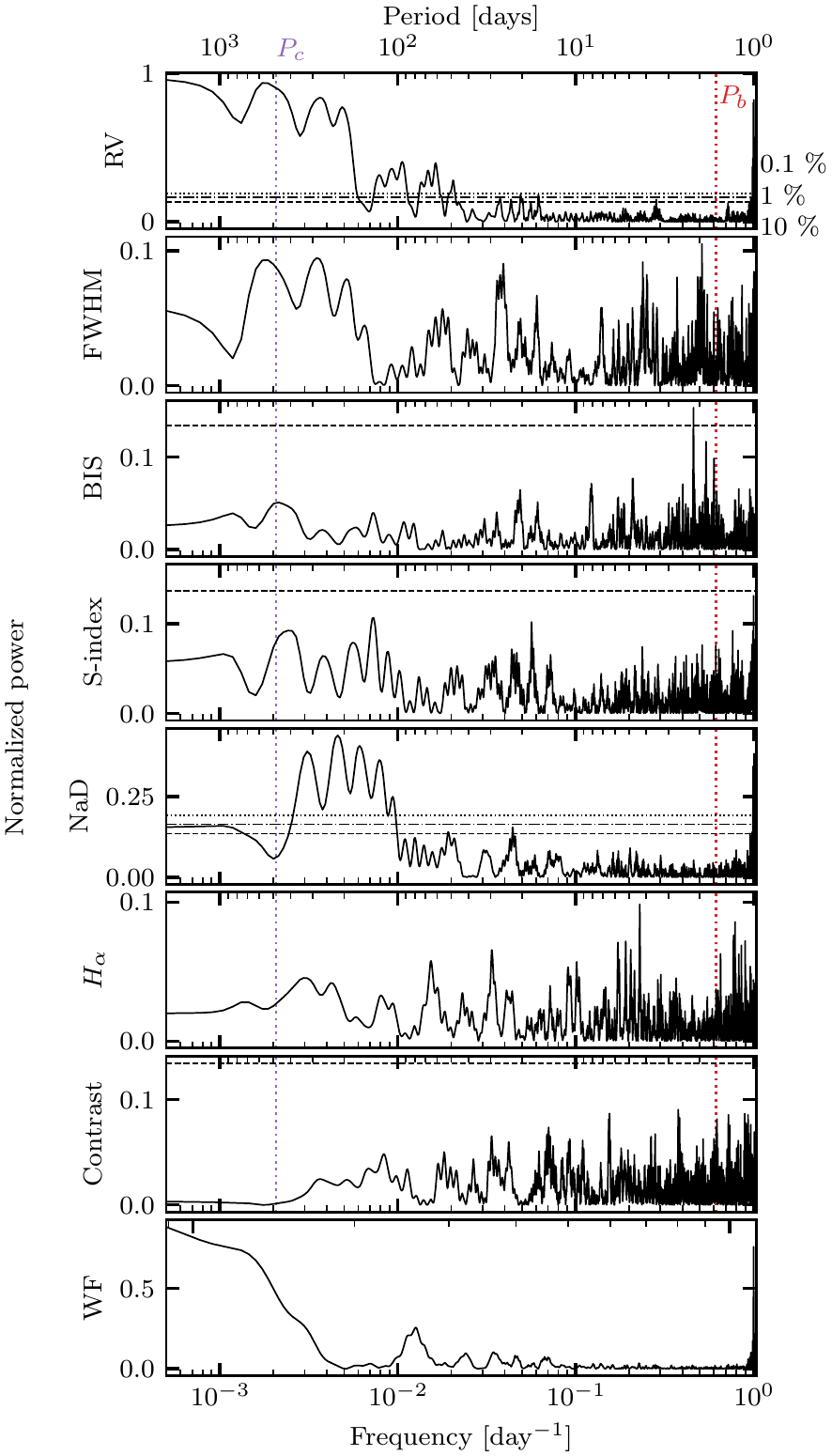}
        \caption{GLS periodograms of the stellar activity indicators from HARPS RVs.
}
    \label{fig:plot_fwhm_bis_si_HARPS}
\end{figure}

From the HARPS and CORALIE data (Table \ref{table:measur_rvs}), we evaluated activity indicators: FWHM, BIS, contrast of the CCF, $H_{\alpha}$ emission, Na-doublet and  the Mount Wilson S-index (HARPS) or directly the Calcium H and K lines (CORALIE). The GLSPs of these indicators are plotted in Figs. \ref{fig:plot_fwhm_bis_si_HARPS} and \ref{fig:plot_fwhm_bis_si_CORALIE} with a logarithmic scale in frequency and in Appendix \ref{Appendix:Linear_GLS} with a linear scale. No significant peak is observed in the GLSP of the activity indicators at the period of the suspected planet b or c\footnote{The period of both planets and the RV offset between the instruments were all fit in the Markov chain Monte Carlo (MCMC) process described on Section \ref{sec:joint}.}.
We also evaluated the correlation between these indicators and the radial velocities from HARPS through the Bayesian framework presented in \cite{2016OLEB...46..385F}\footnote{The code is available in the git repository: \url{https://bitbucket.org/pedrofigueira/bayesiancorrelation/}.}. No significant correlation was observed between the RVs and the BIS, FWHM, or $S_{HK}$ 
(Pearson correlation coefficient of $\rho_{FWHM} = 0.15 \pm 0.08$; $\rho_{BIS} = 0.04 \pm 0.08$; $\rho_{S_{HK}} = 0.01 \pm 0.08$).

\begin{figure}[!htb]
        \centering
        \includegraphics[]{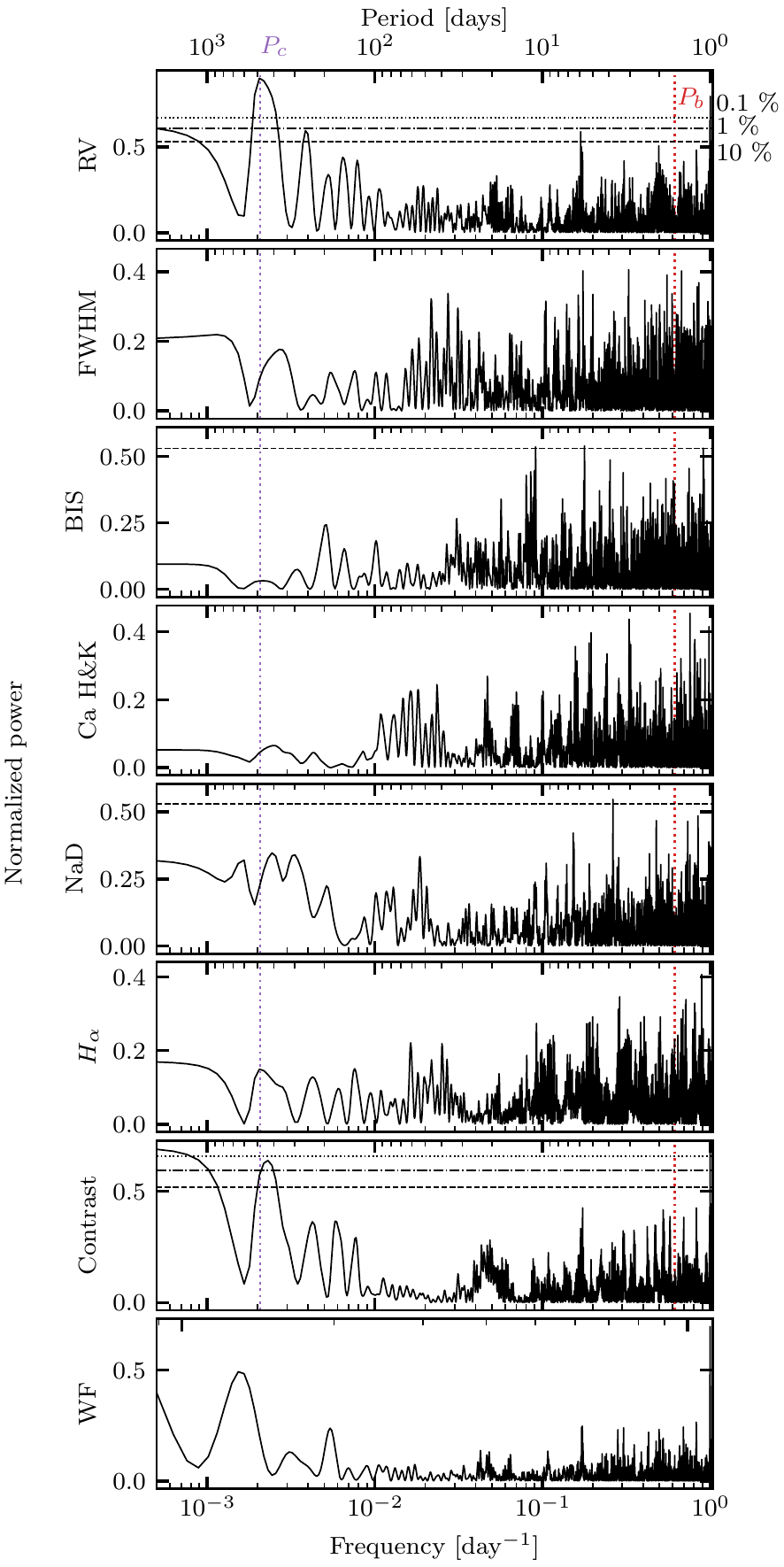}
        \caption{GLS periodograms of the stellar activity indicators from CORALIE RVs.
}
    \label{fig:plot_fwhm_bis_si_CORALIE}
\end{figure}

\section{Stellar characterization} \label{section:stellar}

For the determination of the stellar parameters, we have used the high-resolution, high S/N, HARPS spectra. The individual HARPS spectra were corrected from the RV and co-added in order to produce a unique spectrum with a high S/N. The spectroscopic stellar parameters were derived using the commonly used equivalent width (EW) method ``ARES+MOOG'' (we used ARES v2 and MOOG2014; see \citealt{2014dapb.book..297S} for details). The spectral analysis is based on the excitation and ionization balance of iron abundance. The equivalent widths (EWs) of the lines were consistently measured with the ARES code \citep[][]{Sousa-2007, Sousa-2015b} and the abundances derived in local thermodynamic equilibrium (LTE) with the MOOG code \citep[][]{Sneden-1973}. A grid of plane-parallel Kurucz ATLAS9 model atmospheres was used \citep[][]{Kurucz-1993}. The line list used in this analysis was the same as that used in \citet[][]{Sousa-2008}. This is the same method applied to derive homogeneous spectroscopic parameters for the Sweet-CAT catalog \citep[][]{Santos-2013, Sousa-2018}.

In order to  have accurate estimates of the planetary mass and radius, it is critical to determine those same parameters accurately for the host star. By using the parameters already determined by the spectroscopy analysis as inputs, we then determine the mass, $M_{*}$, radius, $R_{*}$, age, and extinction of HD\,137496 through the use of the Bayesian tool \texttt{PARAM} v1.5 \citep{2017MNRAS.467.1433R}. This pipeline employs Bayesian methods to compute the probability density functions of stellar fundamental properties through the use of a pre-computed grid of stellar evolutionary tracks. The derived stellar parameters are presented in Table \ref{table:stellar}. We then performed independent stellar parameter estimations with different methods using the spectral energy distribution (SED) and isochrones (for more information on these methods, see Appendix \ref{Appendix:Ages}). We found that the mass and radius estimates were compatible across different tools within the 1-$\sigma$ uncertainties. The ages were also compatible within the 1-$\sigma$ uncertainties, with the exceptions of NextGen (see Appendix \ref{Sub-appendix:SED fitting}) and the chemical clocks (see Appendix \ref{Sub-appendix:Chemical clocks}). However, these two methods rely on empirical relations, contrary to all the other methods (including PARAM), which rely on stellar models. The fact that all the methods based on stellar models provide similar estimates despite the use of different stellar evolutionary tracks strengthens our choice to adopt the mass, radius, and age derived with PARAM.

Finally, we derived stellar abundances of several elements using the same tools and models as for the stellar spectroscopic parameter determination, as well as using the classical curve-of-growth analysis method assuming local thermodynamic equilibrium. Although the EWs of the spectral lines were automatically measured with ARES, for the elements with only two-three lines available, we performed a careful visual inspection of the EW measurements. For the derivation of chemical abundances of refractory elements, we closely followed the methods described in \citet[][]{Adibekyan-12, Adibekyan-15, Delgado-14, Delgado-17}. Abundances of the volatile elements, C and O, were derived following the method of \cite{Delgado-10} and \cite{Bertrandelis-15}. Since the two spectral lines of oxygen are usually weak and the 6300.3\AA{} line is blended with Ni and CN lines, the EWs of these lines were manually measured with the task \texttt{splot} in IRAF. Lithium and sulfur abundances were derived by performing spectral synthesis with MOOG following the works by \citet[][]{Delgado-14} and \citet[][]{CostaSilva-20}, respectively. All the [X/H] ratios are obtained by doing a differential analysis with respect to a high S/N solar (reflected from Vesta) spectrum from HARPS. The stellar abundances of the elements are presented in Table \ref{table:abund}. 

We find that the [X/Fe] ratios of most elements are close to solar, as expected for a star with near-solar metallicity. Moreover, we used the chemical abundances of some elements to derive ages through the so-called chemical clocks \citep[i.e., certain chemical abundance ratios that have a strong correlation with age: e.g.,][]{2015A&A...579A..52N}. For more details regarding this estimation, we advise the reader to consult Appendix \ref{Sub-appendix:Chemical clocks}.

The kinematics of the star can also reveal important information about the primordial place where this system was born in the Galaxy. Through the use of \textit{Gaia}'s accurate distance and radial velocity measurements, we were able to calculate the 3D velocities $U$, $V,$ and $W$\footnote{\textit{U}, \textit{V,} and \textit{W} are, respectively, the positive in the directions of Galactic center, Galactic rotation, and the North Galactic Pole.}. These velocities were then adjusted to the space velocities of the Local Standard of Rest (LSR) using the recent LAMOST data \citep{2015ApJ...809..145T}: $U_\odot =9.58$ km/s, $V_\odot = 10.52$ km/s, and $W_\odot = 7.01$ km/s. 
We then calculated the probability of this star belonging to specific populations, the thick disk (TD), thin disk (D), or the stellar halo (H), according to the classical method of \cite{2003A&A...410..527B}. For HD\,137496, we find a TD/D = 0.01, making it clear that this star belongs to the thin disk.

{\renewcommand{\arraystretch}{1.2}

\begin{table}
\caption{\label{table:stellar}IDs, coordinates, and magnitudes of the planet-host star.}
\centering
\begin{tabular}{lcc}
\hline\hline
Parameter&Value&Source\\

\hline

Identifying Information\\

\hspace{3mm} R.A. (J2000) & $15^h 26^m 58^s .07$ & Gaia DR2\tablefoottext{a}\! \\
\hspace{3mm} DEC (J2000) & $-16^h 30^m 31^s .69$ & Gaia DR2\tablefoottext{a}\! \\

\hspace{3mm} HD & 137496 & \\
\hspace{3mm} K2 ID & 364 & \\
\hspace{3mm} EPIC ID & 249910734 & \\

\hline

Distance and Velocities \\

\hspace{3mm} $PM_{R.A.}$ ($mas yr^{-1}$) & 20.54 $\pm$ \, 0.1 & Gaia DR2\tablefoottext{a}\! \\
\hspace{3mm} $PM_{DEC}$ ($mas yr^{-1}$) & -46.86 $\pm$ \, 0.08 & Gaia DR2\tablefoottext{a}\! \\
\hspace{3mm} $\pi$, Parallax (mas) &  6.41 $\pm$ \; 0.07  & Gaia DR2\tablefoottext{a}\! \\
\hspace{3mm} $v_{rad}$ (\kms) &  -2.65 $\pm$ \; 0.34  & Gaia DR2\tablefoottext{a}\! \\
\hspace{3mm} $d$, Distance (pc) & 156 $\pm$ \; 14 & This work \\
\hspace{3mm} $U$ (\kms) &  -10.35 $\pm$ \; 0.31  & This work \\
\hspace{3mm} $V$ (\kms) &  -24.34 $\pm$ \; 0.60  & This work \\
\hspace{3mm} $W$ (\kms) &  -7.89 $\pm$ \; 0.35  & This work \\

\hline

Magnitudes \\

\hspace{3mm} $K_p$ (mag) &  9.833 & MAST EPIC\tablefoottext{b} \\
\hspace{3mm} $G$ (mag) &  9.77 $\pm$ \; 0.34  & Gaia DR2\tablefoottext{a} \\
\hspace{3mm} $B_p$ (mag) &  10.16 $\pm$ \; 0.34  & Gaia DR2\tablefoottext{a} \\
\hspace{3mm} $R_p$ (mag) &  9.25 $\pm$ \; 0.34  & Gaia DR2\tablefoottext{a} \\
\hspace{3mm} $B$ (mag) &  10.63 $\pm$ \; 0.024  & APASS\tablefoottext{c} \\
\hspace{3mm} $V$ (mag) &  9.996 $\pm$ \; 0.028  & APASS\tablefoottext{c} \\
\hspace{3mm} $J$ (mag) &  8.646 $\pm$ \; 0.024  & 2MASS\tablefoottext{d} \\
\hspace{3mm} $H$ (mag) &  8.366 $\pm$ \; 0.046  & 2MASS\tablefoottext{d} \\
\hspace{3mm} $K$ (mag) &  8.244 $\pm$ \; 0.036  & 2MASS\tablefoottext{d} \\
\hspace{3mm} $A_V$ (mag) &  0.211 $\pm$ \; 0.067  & This work \\

\hline

Stellar Parameters \\

\hspace{3mm} $T_\textrm{eff}$ (K) &  5799 $\pm$ 61  & This work \\
\hspace{3mm} log $g$ (cgs) &  4.05 $\pm$ 0.1  & This work \\
\hspace{3mm} [Fe/H] (dex) &  -0.03 $\pm$ 0.04  & This work \\
\hspace{3mm} $v_\textrm{turb}$ (\kms) &  1.16 $\pm$ 0.02  & This work \\
\hspace{3mm} $M_*$\tablefoottext{e} ($M_\odot$) &  1.04 $\pm$ 0.02  & This work \\
\hspace{3mm} $R_*$\tablefoottext{e} ($R_\odot$) &  1.59 $\pm$ 0.03  & This work \\
\hspace{3mm} $\rho_*$ ($\mathrm{g \, cm^{-3}}$) &  0.259 $\pm$ 0.014  & This work \\
\hspace{3mm} Age\tablefoottext{e}  (Gyr) &  8.3 $\pm$ 0.7  & This work \\

\hline

\end{tabular}
\tablefoot{
\tablefoottext{a}{ \cite{2018A&A...616A...1G} } \\
\tablefoottext{b}{ \url{http://archive.stsci.edu/k2/epic/search.php}} \\
\tablefoottext{c}{\cite{2014AJ....148...81M}} \\
\tablefoottext{d}{\cite{2003yCat.2246....0C}} \\
\tablefoottext{e}{We provide here the estimates obtained with PARAM, but we also derived these quantities using other tools (see Section \ref{section:stellar} and Appendix \ref{Appendix:Ages}).} 
}
\end{table}}

{\renewcommand{\arraystretch}{1.2}

\begin{table}
\caption{\label{table:abund}Stellar abundances.}
\centering
\begin{tabular}{lcc}
\hline\hline
Species&Value\\

\hline

Abundances \\

\hspace{3mm} \text{[C I/H]} &    -0.104 $\pm$ \;        0.017 \\
\hspace{3mm} \text{[O I/H]} &    0.045 $\pm$ \; 0.077 \\
\hspace{3mm} \text{[Na I/H]} &   -0.064 $\pm$ \;        0.013 \\
\hspace{3mm} \text{[Mg I/H]} &   0.008 $\pm$ \; 0.016 \\
\hspace{3mm} \text{[Al I/H]} &   0.018 $\pm$ \; 0.015 \\
\hspace{3mm} \text{[Si I/H]} &   -0.029 $\pm$ \;        0.025 \\
\hspace{3mm} \text{[Ca I/H]} &   -0.016 $\pm$ \;        0.041 \\
\hspace{3mm} \text{[Ti I/H]} &   -0.002 $\pm$ \;        0.022 \\
\hspace{3mm} \text{[Cr I/H]} &  -0.043 $\pm$ \; 0.029 \\
\hspace{3mm} \text{[Ni I/H]} &  -0.080 $\pm$ \; 0.014 \\
\hspace{3mm} \text{[Cu I/H]} &  -0.065 $\pm$ \; 0.027 \\
\hspace{3mm} \text{[Zn I/H]} &  -0.076 $\pm$ \; 0.020 \\
\hspace{3mm} \text{[Sr I/H]} &  -0.172 $\pm$ \; 0.077 \\
\hspace{3mm} \text{[Y I/H]} &   -0.134 $\pm$ \; 0.048 \\
\hspace{3mm} \text{[Zr I/H]} &  -0.143 $\pm$ \; 0.036 \\
\hspace{3mm} \text{[Ba I/H]} &  -0.081 $\pm$ \; 0.025 \\
\hspace{3mm} \text{[Ce I/H]} &  -0.058 $\pm$ \; 0.040 \\
\hspace{3mm} \text{[Nd I/H]} &  -0.063 $\pm$ \; 0.034 \\
\hspace{3mm} \text{[S I/H]} &   -0.16 $\pm$ \;  0.05  \\
\hspace{3mm} \text{A(Li)} &     2.02 $\pm$ \;   0.04  \\
\hline

\end{tabular}

\end{table}}

\section{Joint analysis: RV \& photometry} \label{sec:joint}

In order to estimate the planetary parameters, we performed a joint analysis by simultaneously fitting the already normalized \textit{K2} light curve and the RVs from HARPS and CORALIE, using the approach and tools described in \citet{demangeon2018}. A Markov chain Monte Carlo (MCMC) algorithm, implemented in the \texttt{emcee} Python package \citep{2013PASP..125..306F}, was used to sample the posterior probability density function.

To generate the model light curves that are compared to the \textit{K2} light curve, we used the \texttt{batman}\footnote{Available at \url{https://github.com/lkreidberg/batman}.} Python package \citep{2015PASP..127.1161K}. Since the observations from \textit{K2} were taken in long-cadence mode (30-min), we super-sampled\footnote{Integration times can cause morphological distortions to the transit light curve, since we are using long-cadence observations this effect is amplified. Super-sampling is a correction that takes into account the extended exposure times, with the factor of 10 being widely used for K2 and Kepler. This number arises from a comparison with the work of \cite{2010MNRAS.408.1758K}, where the authors studied three targets of Kepler with similar magnitudes and computed that an oversampling factor of 5 is sufficient. In order to avoid any issue regarding the accuracy of our model, due to subtle differences between the targets and HD\,137496, we chose an oversampling factor twice as large (10), which is thus more than enough in terms of accuracy while still manageable in terms of computation time.} the model by a factor of 10. 

To compute the model of the RVs to be compared with the HARPS and CORALIE data, we used the RadVel\footnote{Available at \url{https://github.com/California-Planet-Search/radvel}.} Python package \citep{2018PASP..130d4504F}. Since we are using data from two different spectographs, the model includes an RV offset between the two instruments on top of the systemic velocity, $\gamma_{RV,HARPS}$, as measured by HARPS.

For the two planets, we estimated the period $P$, the time of inferior conjunction $t_{\textrm{ic}}$, the RV semi-amplitude $K$, and the projected eccentricity components, $e \cos \omega $ and $e \sin \omega,$ where $e$ and $\omega$ are the eccentricity and the argument of periastron, respectively; thus, we improve the exploration of the parameter space for low-eccentricity solutions.
For the transiting planet, we also estimated the radius ratio $R_p/R_*$ and the cosine of the orbital inclination $\cos i_{\textrm{p}}$. Finally, the transit signal also enables the derivation of the stellar density $\rho_*$.

To account for sources of noise that would have been ignored during the computation of the error bars provided with the data, we used a jitter parameter added in quadrature to the error bar for the standard deviation of the Gaussian likelihood function \citep[e.g., ][]{baluev2009}. We used one jitter parameter per dataset (\textit{K2}, HARPS, CORALIE).
The priors used are provided in Table \ref{table:results_mcmc}, and more information regarding the reasons behind the choice of priors can be found in Appendix \ref{Appendix:MCMC_priors}.

In order to quicken the convergence despite the use of broad priors, we drew the initial values of the MCMC chains for the planetary parameters $K$, $P$, $t_{ic}$, the cosine of the inclination of planet b, and $\rho_*$ from the posterior distribution of a first fit realized with tighter priors (see Appendix \ref{Appendix:Early_run}).
Before running the \texttt{emcee} MCMC exploration, we carried out a pre-optimization of the parameter's values using the downhill simplex (aka amoeba) algorithm. 
We performed one optimization for each walker and used the results as initial values of the \texttt{emcee} exploration. The number of walkers was defined as three times the number of free parameters rounded to the closest even number. We computed 100,000 steps for each Markov chain walker. After the run, we removed the burn-in from the chains and studied the convergence using a modified Geweke diagnostic \citep{geweke1991evaluating}. The planetary parameters were then estimated by the median of the resulting marginalized posterior distributions and the borders of the 68\,\% confidence interval as the 16$^{th}$ and 84$^{th}$ percentiles.

\begin{figure*}[!htb]
    \centering
    \subfloat{\includegraphics[width = 0.372\linewidth]{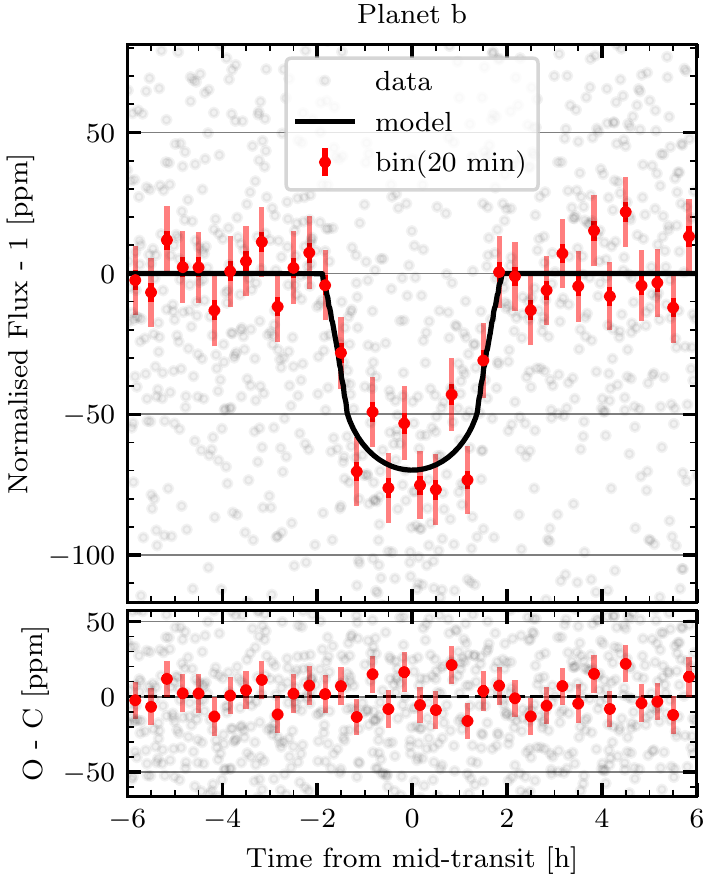}\label{fig:a}}
    \hspace{0.5cm}
    \subfloat{\includegraphics[width = 0.57\linewidth]{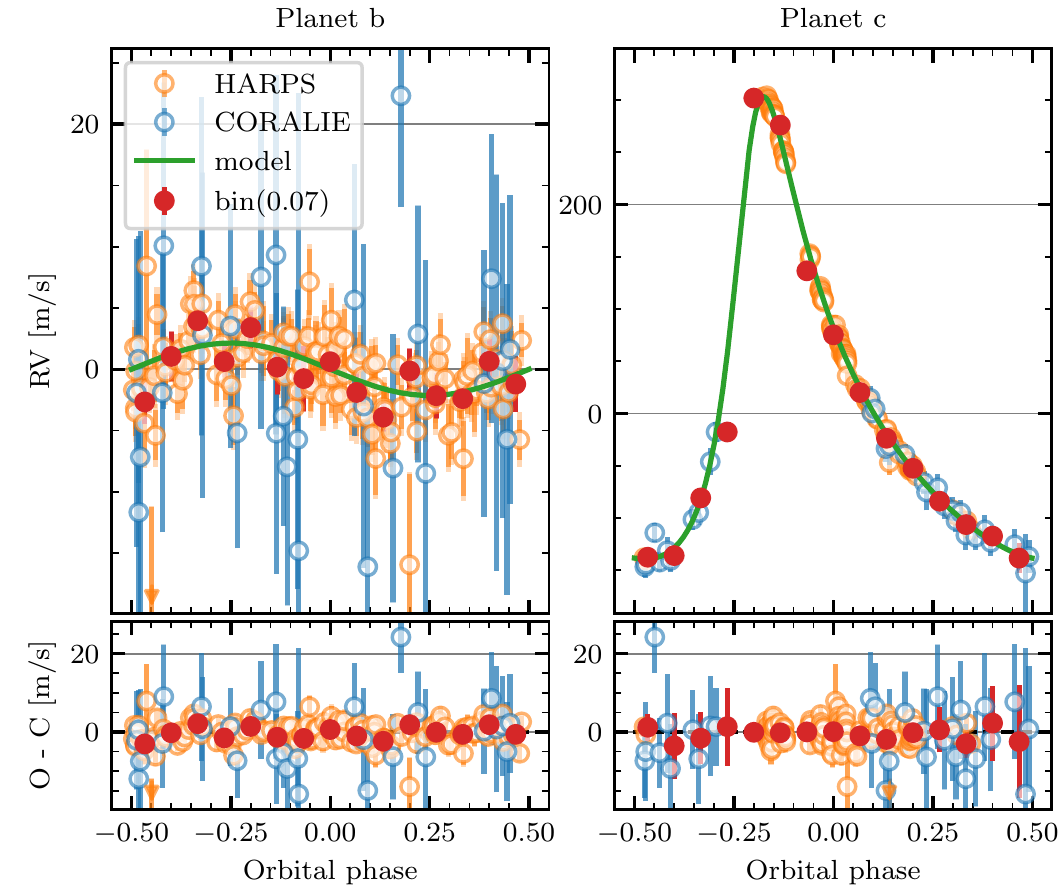}\label{fig:b}}
    \caption{Left: Phase-folded and binned transit with residuals for HD\,137496\,b together with the best fitting model. Root mean square (rms) of the K2 residuals: 71 ppm and 10 ppm for bins in phase corresponding to 20 min. Right: Phase-folded RV fit with residuals for HD\,137496\,b (left) and HD\,137496\,c (right). Rms of ths HARPS residuals: 3.4 m/s; CORALIE residuals: 6.6 m/s.}
    \label{fig:final_phases}
\end{figure*}

The full list of fitted parameters along with the obtained values and confidence intervals are shown in Table \ref{table:results_mcmc}. The phase-folded light curve and RVs and the best-fit model using the estimated parameters are shown in Fig. \ref{fig:final_phases}.

{\renewcommand{\arraystretch}{1.5}

\begin{table*}[h!]
\caption{\label{table:results_mcmc}Instrumental, stellar, and planetary parameters derived from the joint analysis on the photometric and radial velocity data.}
\centering 
\begin{tabular}{lcc}
\hline\hline
Parameter & Results - This work & Prior\tablefoottext{a}\!\\

\hline

Instrument parameters \\

\hspace{3mm} Systemic velocity $\gamma_{RV,HARPS}$ [\ms] & $-2988.9_{-1.0}^{+1.1}$ &  $\mathcal{G}$ [-2921.8, 130] \\
\hspace{3mm} \textit{K2} jitter [ppm] & $66.8_{-1.38}^{+1.43}$ &  $\mathcal{J}$ [$1.9\cdot 10^{-6}$, $1.9 \cdot 10^{-4}$]\tablefoottext{b}\! \\
\hspace{3mm} HARPS radial velocity jitter [\ms] & $1.52_{-0.23}^{+0.23}$ & $\mathcal{J}$ [0.017, 17]\tablefoottext{b}\! \\
\hspace{3mm} CORALIE radial velocity jitter [\ms] & $0.72_{-0.51}^{+1.88}$ & $\mathcal{J}$ [0.12, 120]\tablefoottext{b}\! \\
\hspace{3mm} $\Delta \text{RV}$ CORALIE to HARPS  [\ms] & $-18.64_{-2.46}^{+2.48}$ & $\mathcal{U}$ [-111, 11]\\

\hline
Stellar parameters \\

\hspace{3mm} Stellar density $\rho_*$ [$\textrm{g.cm}^{-3}$] & $0.259_{-0.014}^{+0.014}$ &  $\mathcal{G}$ [0.2590, 0.0150] \\
\hspace{3mm} Limb-darkening coefficient $u1$
 & $0.5325_{-0.0061}^{+0.0061}$ & $\mathcal{G}$ [0.5324, 0.0061]\\
\hspace{3mm} Limb-darkening coefficient $u2$ & $0.1056_{-0.0089}^{+0.0090}$ & $\mathcal{G}$ [0.1056, 0.0091]\\

\hline

\multicolumn{3}{c}{\textbf{HD\,137496\,b}}  \\
\hline

\hspace{3mm} Period $P_b$ [days] & $1.62116_{-8.06  \, 10^{-5}}^{+7.91\, 10^{-5}}$ & $\mathcal{J}$ [0.09, 1400] \\
\hspace{3mm} RV semi-amplitude $K_b$ [\ms] & $2.14 \pm 0.29$ & $\mathcal{U}$ [0.0, 50.0] \\
\hspace{3mm} Radius ratio $R_{p,b}/R_*$  & $0.0076_{-0.00029}^{+0.00030}$ & $\mathcal{U}$ [0.0, 1.0] \\
\hspace{3mm} Impact parameter $b_b$  & $0.27_{-0.17}^{+0.14}$ & $\mathcal{U}$ [0.0, 2.0] \\
\hspace{3mm} Eccentricity $e_b$  & 0 & Fixed at 0\!\\
\hspace{3mm} Time of inferior conjunction $t_{ic,b}$ [\text{BJD{\text -}2450000}] & $8039.1317_{-0.0029}^{+0.0028}$ & $\phi$ : $\mathcal{U}(0, 1)$\tablefoottext{c}\!\\
\hspace{3mm} Transit depth [ppm] & $57.5_{-4.28}^{+4.61}$ & - \\
\hspace{3mm} Transit duration $T_b$ [hr] & $3.28_{-0.14}^{+0.10}$ & - \\
\hspace{3mm} Ingress to egress duration $\tau_b$ [hr] & $3.23_{-0.15}^{+0.11}$ & - \\
\hspace{3mm} Scaled semi-major axis $a_b/R_*$  & $3.70  \pm 0.07$ & - \\
\hspace{3mm} Semi-major axis $a_b$ [AU]  & $0.02732  \pm 0.00019$ & - \\
\hspace{3mm} Inclination $i_b$ [º] & $85.8_{-2.2}^{+2.6}$ & - \\
\hspace{3mm} Radius $R_b$ [$R_\oplus$] & $1.31_{-0.05}^{+0.06}$ & - \\
\hspace{3mm} Mass $M_b$ [$M_\oplus$] & $4.04 \pm 0.55$ & - \\
\hspace{3mm} Density $\rho_b$ [$\mathrm{g \, cm^{-3}}$] & $10.49_{-1.82}^{+2.08}$ & - \\
\hspace{3mm} Equilibrium temperature $T_{eq,b}$ [K] & $2130_{-29}^{+30}$ & - \\

\hline

\multicolumn{3}{c}{\textbf{HD\,137496\,c}}  \\
\hline

\hspace{3mm} Period $P_c$ [days] & $479.9_{-1.0}^{+1.1}$ & $\mathcal{J}$ [0.09, 1400] \\
\hspace{3mm} RV semi-amplitude $K_c$ [\ms] & $221.25_{-0.82}^{+0.83}$ & $\mathcal{U}$ [0, 444] \\
\hspace{3mm} Eccentricity $e_c$  & $0.477 \pm 0.004$ & $\mathcal{B}$ [0.867, 3.03] \\
\hspace{3mm} Time of inferior conjunction $t_{ic, c}$ [\text{BJD{\text -}2450000}] & $8629.21 \pm 0.41$ & $\phi$ : $\mathcal{U}(0, 1)$\tablefoottext{c}\!\\
\hspace{3mm} Argument of periastron [deg] $\omega_c$  & $-39.01_{-0.47}^{+0.46}$ & - \\
\hspace{3mm} Semi-major axis $a_c$ [AU]  & $1.2163_{-0.0088}^{+0.0087}$ & - \\
\hspace{3mm} Minimum mass $M_c \sin i_c$ [$M_{Jup}$] & $7.66 \pm 0.11$ & - \\
\hspace{3mm} Equilibrium Temperature $T_{eq,c}$ [K] & $370 \pm 5$ & - \\

\hline

\end{tabular}

\tablefoot{In the column on the right, we display our prior choices: $\mathcal{U}$ - Uniform; $\mathcal{G}$ - Gaussian; $\mathcal{J}$ - Jeffreys; $\mathcal{B}$ -  beta distribution. \\
\tablefoottext{a}{For more information, see Appendix \ref{Appendix:MCMC_priors}.} \\
\tablefoottext{b}{The Gaussian prior distribution is truncated to only consider positive values.}\\
\tablefoottext{c}{We define $\phi$ the orbital phase of the time of inferior conjunction as $\phi$ = ($t_{ic}$ - $t_{ref}$) / $P$,  where $t_{ref}$ is an arbitrarily chosen reference time. We then considered a uniform prior between 0 and 1 to $\phi$. We chose $t_{ref}$ to be 8039 and 8620 BJD-2450000 for planets b and c, respectively.}
}

\end{table*}}

\section{Discussion} \label{sec:conc}

In this work, we present the HD\,137496 system, composed of two planets orbiting a G-type star near the main-sequence turn-off. To characterize the two planets, we acquired high precision radial velocity observations with the HARPS and CORALIE spectrographs together with \textit{K2} photometry which exhibits transits of HD\,137496\,b. We present a mass measurement for HD\,137496\,b of $M = 4.04 \pm 0.55$ $M_\oplus$ with a radius of $R = 1.31_{-0.05}^{+0.06}$ $R_\oplus$ and a density of $\rho$ = $10.49_{-1.82}^{+2.08}$ $\mathrm{g \, cm^{-3}}$, while the non-transiting HD\,137496\,c has a minimum mass of  $7.66 \pm 0.11$ $M_{\textrm{Jup}}$.

\subsection{A low-mass inner planet with a cold distant companion} 

This system is composed of an eccentric cold Jupiter and a hot super-Mercury (or dense super-Earth). This planetary architecture has gained traction in the exoplanet field in recent years \citep{2016A&A...592A..13S, 2017AJ....153..210H, 2018AJ....156...92Z, 2020A&A...643A..66B}. As mentioned in Section \ref{sect:intro},  \cite{2018AJ....156...92Z} found a correlation between the presence of inner super-Earths and long-period Jupiter-like planets with \textit{Kepler} data. The authors derived that cold Jupiters appear three times more often around hosts of super-Earths than they do around field stars, suggesting that cold Jupiters do not directly compete with inner super-Earths for solid material, thus allowing their formation. However, the recent study by \cite{2020A&A...643A..66B} brings to light the eccentricity of the distant massive companion in these statistics. The authors argue that if the giant planet has an eccentric orbit, it is very unlikely that the system harbors any inner super-Earths (making the HD 137496 system a counter example). Nevertheless, the presence of a low-mass inner planet together with a cold Jupiter in eccentric and/or mutually inclined orbits has been the object of recent studies \citep{2017MNRAS.467.1531H, 2019AJ....157..145M, 2019A&A...629L...7C, 2020AJ....159...38M, 2020MNRAS.498.5166P, 2021MNRAS.508..597P}. This highlights the need for accurate characterization of such systems for our understanding of planet formation and evolution and stresses the importance of the HD\,137496 system.

The high eccentricity of the orbit of HD\,137496\,c ($e \simeq 0.48$) is expected for high-mass planets \citep{2007A&A...464..779R, 2013A&A...560A..51A}, hinting at a dynamically "hot"\footnote{By dynamically “hot” we mean that strong planet-planet dynamic interactions caused significant changes to their orbital parameters.} evolution \citep{2013A&A...555A.124B, 2013ApJ...775...42I, 2020arXiv200205756R}. In light of these recent works, this could be interpreted as the effect of formation and migration mechanisms for which the orbit of HD\,137496\,b was tidally circularized at the currently observed short orbital period. 

\subsection{Composition of HD\,137496\,b}

\begin{figure}[!htb]
        \centering
        \includegraphics[width=\linewidth]{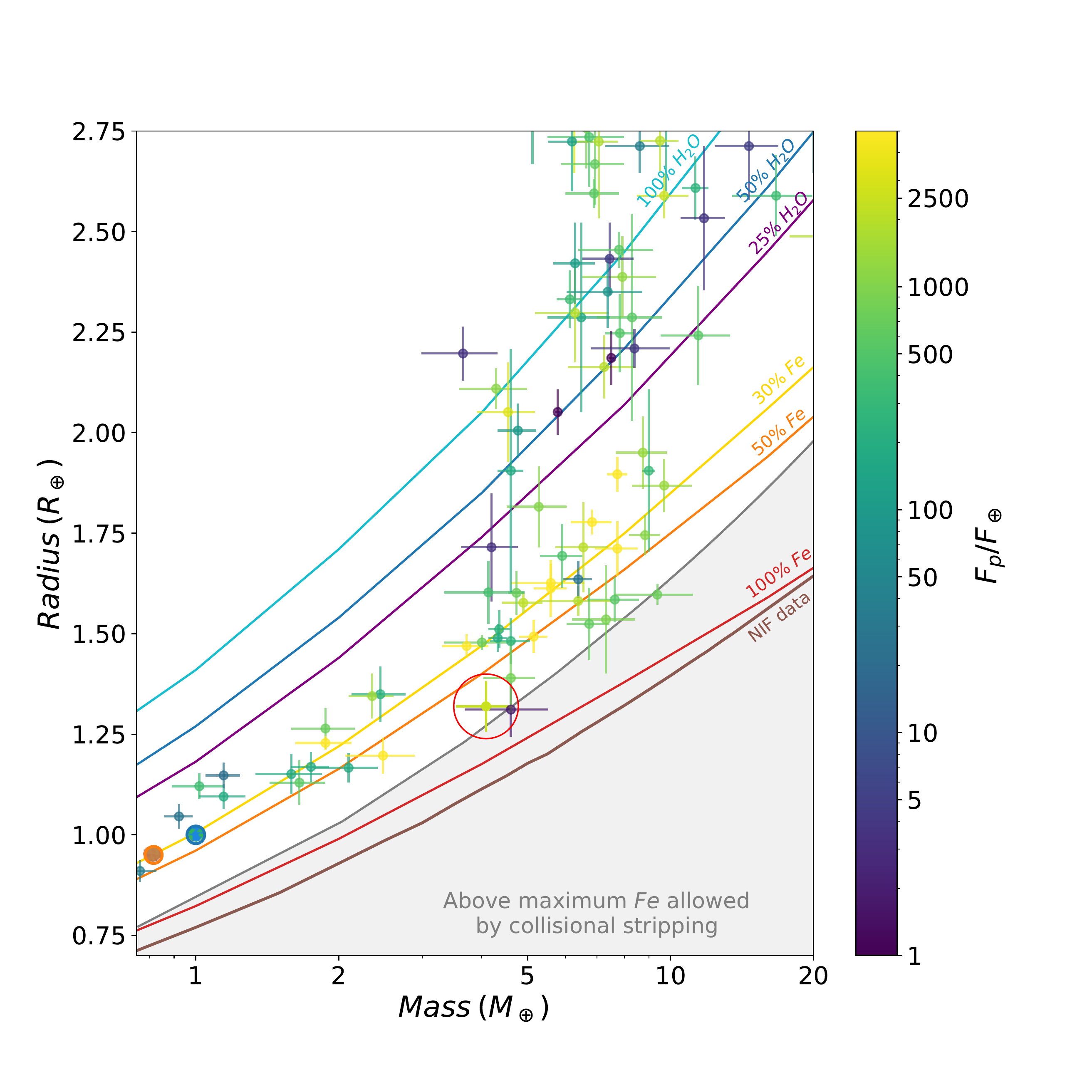}
        \caption{Mass–radius diagram for planets with radii below 5 $R_\oplus$ and masses below 20 $M_\oplus$, with mass uncertainties under 20\%. Planets are color-coded according to their incident flux. Data retrieved from TEPcat \citep{2011MNRAS.417.2166S} as of March 2021. For reference, Venus and Earth are also plotted. 
Colored lines represent the theoretical models for internal composition of small planets \citep{2016ApJ...819..127Z}. The experiment-based mass-radius relationship for a hypothetical pure iron planet from \cite{2018NatAs...2..452S} is shown in brown, with data from the National Ignition Facility (NIF). The gray line and shaded area represent the limit by maximum collisional stripping curve according to \cite{2010ApJ...712L..73M}. The small planet discovered and characterized in this paper is plotted in the red circle.
}
    \label{fig:tepcat}
\end{figure}

In Fig. \ref{fig:tepcat}, we show HD\,137496\,b in the mass-radius diagram alongside all the small planets ($M < 20 M_\oplus$)  extracted from TEPcat\footnote{Available at \url{http://
www.astro.keele.ac.uk/jkt/tepcat/}.} \citep{2011MNRAS.417.2166S} whose masses are known with a precision better than 20\%. According to the models of \cite{2016ApJ...819..127Z}, HD\,137496\,b has the bulk density of a planet composed mostly of iron (consistent within the uncertainties). In the literature, very few planets appear to have densities as high as HD\,137496\,b, making it a reference and a compelling target for the study of interior models and planet formation. 

\subsubsection{Modeling of the internal structure of HD\,137496\,b}

Using the estimated mass and radius, we computed a marginal core radius fraction (CRF) of 81\% $\pm$ 11\% using hardCORE\footnote{Available at \url{https://github.com/gsuissa/hardCORE/blob/master/hardcore.py}.} \citep{2018MNRAS.476.2613S}. This simple model assumes that the planet is fully differentiated with a maximum density layer no denser than iron. The model then calculates the minimum and maximum core radius fractions for which a marginal CRF can be sampled. The core radius fraction is related \citep{2017ApJ...837..164Z} to the core mass fraction (CMF) by 


\begin{equation}
    CRF \approx \sqrt{CMF}.
\end{equation}
This translates to a CMF of 78\% $\pm$ 11 \% using this model.

 As an alternative method, we characterized the internal structure of HD\,137496\,b considering a pure-iron core, a silicate mantle, and a pure-water layer using a Bayesian inference analysis. We did not add hydrogen and helium to the model since such a dense (10.49$~g$/$cm^3$) and hot ($T_{\textrm{eq}} = 2130$\,K) planet is not expected to retain a significant volatile envelope \citep[e.g.,][]{2007ApJ...669.1279S}. The equations of state (EOS) used for the iron core are taken from \cite{Hakim2018},  the EOS of the silicate-mantle is calculated with PERPLE\_X from \cite{Connolly09} assuming Earth-like abundances, and  we used the AQUA EOS from \cite{2020A&A...643A.105H} for the pure-water layer. The thickness of the planetary layers were set by defining their masses and solving the structure equations. 
  
  \begin{figure*}[h!]
\centering
  \begin{tabular}{@{}cc@{}}
    \includegraphics[scale=0.45]{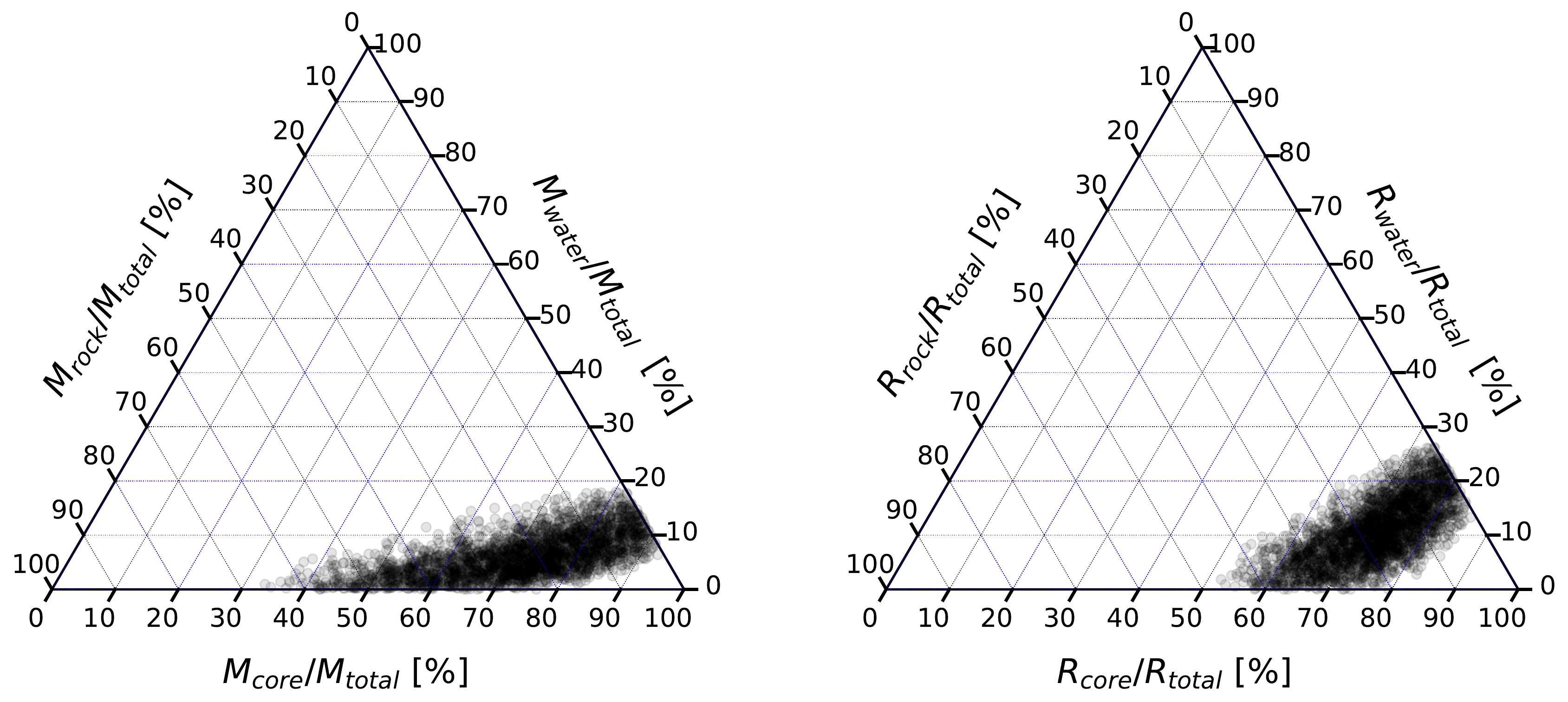}
  \end{tabular}
  \caption{Ternary diagram of inferred internal composition of HD\,137496\,b. We show the parameter space covered by the posterior distributions in mass (left) and radius (right).}  
  \label{fig:ternary}
\end{figure*}

In order to quantify the degeneracy between various interior parameters and produce posterior probability distributions, we used a generalized Bayesian inference analysis with a nested sampling scheme \citep{buchner2014}. Fig. \ref{fig:ternary} shows ternary diagrams of the inferred composition of HD\,137496\,b, which shows the degeneracy linked to the determination of the composition of exoplanets. The degeneracy of internal compositions leading to the observed mass and radius is relatively low since a planet with such a high density needs to have a large iron core. We find an iron mass fraction (IMF) of $0.73 \pm 0.12$, which is in agreement with the estimate from hardCORE. 

The silicon mantle and water layer have estimated relative mass fractions of 23\% and 7\%, respectively. The stellar abundance ratios can be used as a proxy for the bulk abundances of the planet, reducing the degeneracy of internal models leading to the same mass and radius \citep{Dorn_2015, 2021arXiv210212444A}. However, the planet bulk density of HD\,137496\,b is relatively large, while the stellar abundance does not show any particularly iron-rich refractory composition.

In fact, if we assume that the ratios of Fe/Si and Mg/Si measured in the host star reflect the planetary ratios, we are not able to match the observed mass and radius. This is illustrated in Figure \ref{fig:tepcat}, where the planet sits well below the area that represents purely rocky interiors reflecting the host star abundance within 1-$\sigma$.  

\begin{figure}[!htb]
        \centering
        \includegraphics[width=\linewidth]{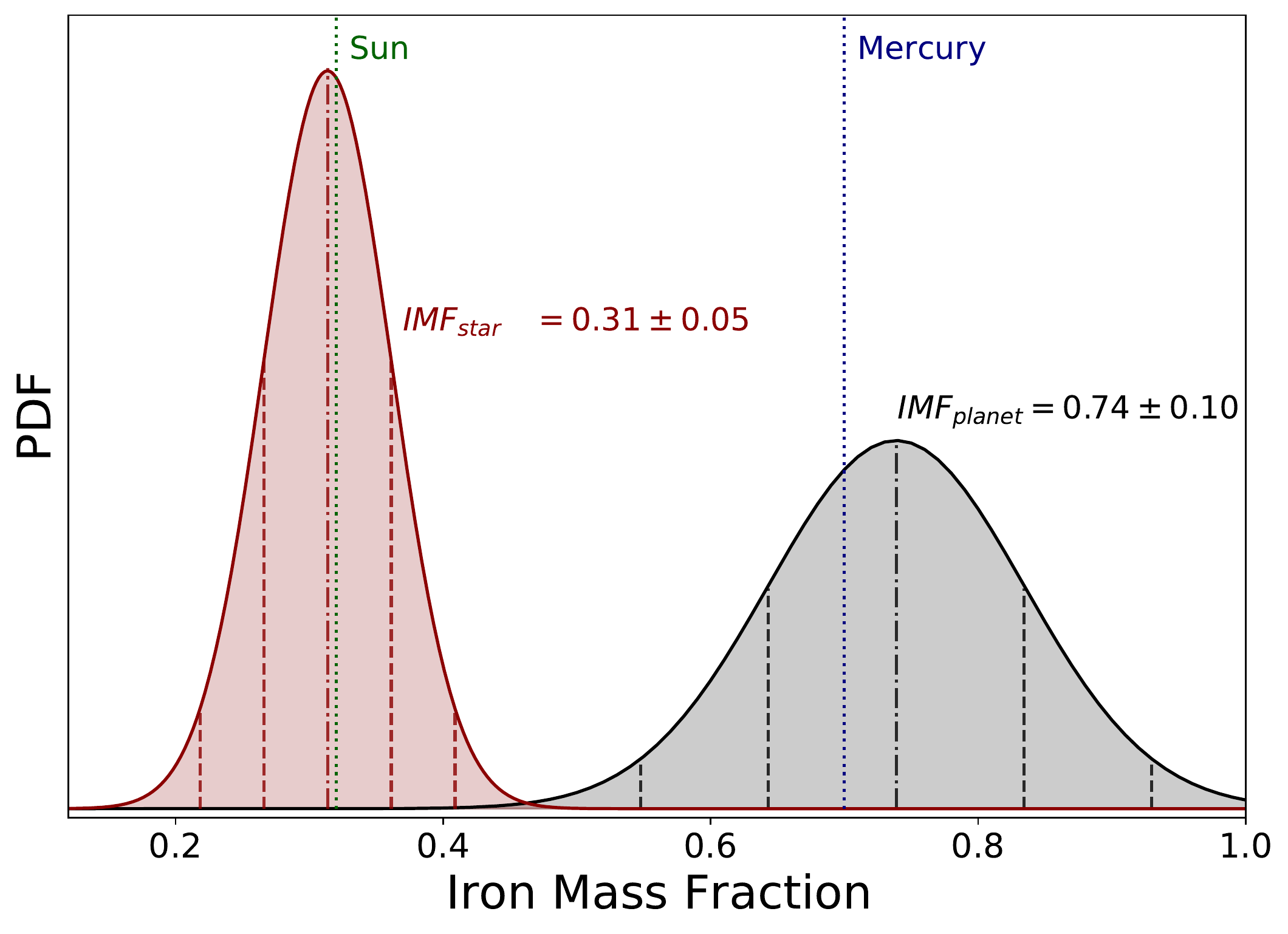}
        \caption{Posterior distribution for estimated iron mass fractions (IMF) as inferred from stellar refractory abundances (red) and the planet mass and radii (gray). All IMF calculations assume that the cores are pure iron and silicate mantles are iron-free. The 2-$\sigma$ regions of both posterior distributions do not overlap, solidifying arguments in favor of a likely super-Mercury interior of HD\,137496\,b with high IMF.
}
    \label{fig:cmf-planet-star}
\end{figure}

To emphasize the unusual high density of HD\,137496\,b, we compare the planet's inferred iron mass fraction (IMF$_{\rm planet}$) to its stellar counterpart (IMF$_{\rm star}$) in Figure \ref{fig:cmf-planet-star}. The IMF$_{\rm star}$ is calculated from [Fe/H], [Si/H], and [Mg/H] (Tables \ref{table:stellar} and \ref{table:abund}), assuming that all Mg and Si are oxidized \citep{2007ApJ...669.1279S, 2010ApJ...712..974R, Dorn_2015, 2016ApJ...819...32U, 2018Icar..299..460W}. Similarly, the calculation of IMF$_{\rm planet}$ assumes that the core is pure iron. Clearly, the IMF$_{\rm planet}$ is similar to Mercury and the 2-$\sigma$ regions of both IMF$_{\rm star}$ and IMF$_{\rm planet}$ do not overlap. This means that the probability of the planet reflecting the stellar refractory abundances is less than 0.04\%.

This high IMF is also beyond the values expected for a star with a solar-like chemical composition. As shown in \cite{2007ApJ...665.1413V}, by assuming a solar nebula composition, we have a [Fe/Si] ratio constrained to $\sim$ 65\%. The star we studied here is similar to the Sun ($M = 1.035 \pm \ 0.022 \; M_\odot$), with abundances derived from the HARPS spectrum shown in Table \ref{table:abund}. From Tables \ref{table:stellar} and \ref{table:abund}, we have [Fe/H] = -0.027 $\pm$ 0.04 and [Si/H] = -0.029 $\pm$ 0.025, resulting in [Fe/Si] = 0.002 $\pm$ 0.027 signaling a Fe/Si close to the one observed in the Sun. 

Similar iron-enriched planets were recently identified \cite[][]{2018NatAs...2..393S, Plotnykov, 2021PSJ.....2..113S}. Among the best characterized planets, there are only a few such super-Mercury planets \citep{otegi2020revisited, 2021arXiv210212444A}, and only a handful for which stellar abundances were measured \citep[e.g., Kepler 107\,c,][]{bonomo2019giant}. HD\,137496\,b is thus a precious object when it comes to understanding the formation of super-Mercuries.


\begin{table}
\centering
\caption{Inferred interior structure properties of HD\,137496\,b.}
\begin{tabular}{lcccc}
\hline
  \ \  \ \ \     $M_{core}/M_{total}$&&&& $0.73^{+0.11}_{-0.12}$\\
  \ \  \ \ \      $M_{mantle}/M_{total}$&&&& $0.21^{+0.16}_{-0.14}$\\
  \ \  \ \       $M_{water}/M_{total}$&&&& $0.06^{+0.05}_{-0.04}$\\
  \hline

\end{tabular}

\end{table}

\subsubsection{Formation of high-density planets}

Considering the collisional history during planet formation, \citet{scora2020chemical} found that it is difficult to form super-Mercuries and that a series of extremely energetic impacts would be needed to form iron-rich planets. Furthermore, such impact events are rare and IMFs are expected to only increase by a factor of 2 in extreme cases.  Within observational uncertainties, this can, in principle, explain the iron enrichment of HD\,137496\,b, but the most common spread in IMF within a system is on the order of only 10\% \citep{scora2020chemical}, which is insufficient to explain the iron enrichment of HD\,137496\,b. Interestingly, by invoking giant impacts, \citet{2010ApJ...712L..73M} argue that the formation of super-Mercuries might be limited to masses below 5 \MEarth, similar to that of HD\,137496\,b. This is because an even larger impactor of more than 10 \MEarth would be required and most planets of such high masses have thick volatile envelopes, which might lead to very different collision outcomes. The special conditions required to form high-density planets is consistent with the fact that these objects are mostly uncommon.

As in the giant impact scenario, favored by a small semi-major axis and their associated high orbital velocity, other processes are enhanced by the close proximity to the host star. As theorized in the studies of Mercury, the close proximity to the host star, and the ensuing high surface temperature of the planet can cause the evaporation of the silicate-rich mantle \citep{1985Icar...64..285C, 2013MNRAS.433.2294P}, increasing the estimated iron-mass fraction. The planet’s environment in the proto-planetary disk can also heavily influence its density \citep{2021arXiv210212444A}. Processes such as photophoresis \citep{2013ApJ...769...78W}, rocklines \citep{2020ApJ...901...97A}, magnetic erosion \citep{2014Icar..241..329H}, magnetic boost \citep{2018ApJ...869...45K, 2020PSJ.....1...23K}, and others can cause the formation of Mercury-like planets in the innermost regions of the disk. 

A combination of all these phenomena may explain the density observed on HD\,137496\,b. Nonetheless, more "exotic" scenarios are also a possibility. An example is that we may be observing a planetary core that is still compressed \citep{2014RSPTA.37230164M}. If a recent (< a few gigayears ago) event like the Roche lobe overflow \citep{ 2014ApJ...793L...3V, 2015ApJ...813..101V, 2016CeMDA.126..227J, 2017ApJ...846L..13K, 2017MNRAS.469..278G}, 
an increased evaporation rate of the planet's gaseous envelope (e.g., caused by an abrupt decrease of the star-planet distance due to orbital tidal decay or migration), or another unknown process that freshly peeled off the atmosphere of a gaseous planet, the ensuing core-remnant would then sustain an exceptionally high density.
Regardless of which mechanisms are at play here, HD\,137496\,b, with well-characterized planetary and stellar parameters, is an ideal target for future works regarding the formation of high-density planets.

\subsection{Future observations}

As observed in Mercury \citep{potter1997sodium, 2015P&SS..115...90P}, this planet may host a mineral atmosphere and is a promising target to search for an elongated comet-like planet’s exosphere. The small distance to the  star, together  with  an  internal  structure  that  resembles  that of Mercury, might create the conditions for an exospheric emission of the heavy refractory elements present at the planet’s surface. Such a tail could extend within several planetary radii and disclose relevant clues regarding the planet’s chemical composition \citep{mura2011comet, merkel2018evidence, 2018MNRAS.481.5296V}. Future transmission spectroscopy observations could bring into sight possible spectral signatures from this planetary tail.
As studied in previous systems \citep[e.g., $\pi$~Men~c - ][]{2020A&A...642A..31D, 2021MNRAS.502.2893K}, measurements of the planetary inclinations and spin-orbit alignment would also help to constrain the formation and evolution models for this system.

\section{Conclusion}

We report the discovery of a planetary system orbiting HD\,137496 (K2-364). With photometry from K2 and precise RVs from both HARPS and CORALIE spanning over two years, we are able accurately characterize the stellar host, study its chemical composition, and put strong constraints on the planetary companions. We found the planetary system to be composed of a cold ($a = 1.22 \pm 0.01$ AU), massive ($M = 7.66 \pm 0.11 M_{Jup}$) giant planet together with an inner ($a = 0.02732 \pm 0.00019$ AU) super-Mercury, for which we achieve a precise determination ($\sim$20\,\% of relative precision) of the bulk density ($\rho$ = $10.49_{-1.82}^{+2.08}$  $\mathrm{g \, cm^{-3}}$; $M = 4.04 \pm 0.55$ $M_\oplus$ and $R = 1.31_{-0.05}^{+0.06}$ $R_\oplus$). Our interior modeling analysis shows that this small planet has an iron-dominated composition with the core representing over 70\% of the planet's mass.

HD\,137496\,b (K2-364\,b) joins the small sample of well-characterized dense planets, making it an interesting target for testing planet formation theories, density enhancing mechanisms, and even the possible presence of an extended comet-like mineral rich exosphere. Together with HD\,137496\,c (K2-364\,c), a high-mass (mass ratio $m_c/m_b$>400), high-eccentricity planet, this system presents an interesting architecture for planetary evolution studies. Future astrometric observations could also provide significant constraints on the relative inclination of the planetary orbits, unraveling new opportunities to discover the system's dynamical history.





\begin{acknowledgements} 

We thank Morgan Deal and Nuno Moedas for helpful discussion on the effects of atomic diffusion on chemical abundances.

This work was supported by Fundação para a Ciência e a Tecnologia (FCT) and Fundo Europeu de Desenvolvimento Regional (FEDER) via COMPETE2020 through the research grants UIDB/04434/2020, UIDP/04434/2020, PTDC/FIS-AST/32113/2017 \& POCI-01-0145-FEDER-032113, PTDC/FIS-AST/28953/2017, PTDC/FIS-AST/3388/2020 \& POCI-01-0145-FEDER-028953. \\

The French group acknowledges financial support from the French Programme National de Plan\'etologie (PNP, INSU). \\

We  thank  the  Swiss  National  Science  Foundation  (SNSF) and the Geneva University for their continuous support to our planet search programs. This work has been in particular carried out in the frame of the National Centre for Competence in Research {\it PlanetS} supported by the Swiss National Science Foundation (SNSF). \\ 
This publication makes use of The Data \& Analysis Center for Exoplanets (DACE), which is a facility based at the University of Geneva (CH) dedicated to extrasolar planets data visualisation, exchange and analysis. DACE is a platform of the Swiss National Centre of Competence in Research (NCCR) PlanetS, federating the Swiss expertise in Exoplanet research. The DACE platform is available at \url{https://dace.unige.ch}. \\ 

This study is based on observations collected at the European Southern Observatory under ESO programme 1102.C-0249. \\ 
This paper includes data collected by the Kepler mission and obtained from the MAST data archive at the Space Telescope Science Institute (STScI). Funding for the Kepler mission is provided by the NASA Science Mission Directorate. STScI is operated by the Association of Universities for Research in Astronomy, Inc., under NASA contract NAS 5–26555.\\
This work has made use of data from the European Space Agency (ESA) mission {\it Gaia} (\url{https://www.cosmos.esa.int/gaia}), processed by the {\it Gaia} Data Processing and Analysis Consortium (DPAC,
\url{https://www.cosmos.esa.int/web/gaia/dpac/consortium}). Funding for the DPAC has been provided by national institutions, in particular the institutions participating in the {\it Gaia} Multilateral Agreement. \\

T.A.S acknowledges support from the Fundação para a Ciência e a Tecnologia (FCT) through the Fellowship PD/BD/150416/2019 and POCH/FSE (EC).

O.D.S.D. is supported in the form of work contract (DL 57/2016/CP1364/CT0004) funded by FCT.

D.J.A. acknowledges support from the STFC via an Ernest Rutherford Fellowship (ST/R00384X/1).

E.D.M. acknowledges the support by the Investigador FCT contract IF/00849/2015/CP1273/CT0003.

V.A. acknowledges the support from FCT through Investigador FCT contract nr.  IF/00650/2015/CP1273/CT0001

L.D.N. is funded by the Swiss National Science Foundation (P2GEP2\_200044).

C.D. acknowledges support from the Swiss National Science Foundation under grant PZ00P2\_174028.

J.L-B. acknowledges financial support received from ”la Caixa” Foundation (ID 100010434) and from the European Union’s Horizon 2020 research and innovation programme under the Marie Skłodowska-Curie grant agreement No 847648, with fellowship code LCF/BQ/PI20/11760023.

DJAB acknowledges support from the UK Space Agency.

A.H. acknowledges support from an STFC studentship.

S.H. acknowledges CNES funding through the grant 837319.

O.M. acknowledges support from CNES.

A.O. acknowledges support from an STFC studenship. 

P.J.W. acknowledges support from STFC consolidated grants (ST/P000495/1 and ST/T000406/1).

\end{acknowledgements}

\bibliographystyle{aa} 
\bibliography{bib_file.bib} 

\appendix

\section{K2 PDCsap photometric data and the reduced data through the POLAR pipeline}

\begin{table}[H]
\caption{Photometric data.}
\label{table:photo}
\centering
\begin{tabular}{ccccccc}
\hline
&
\multicolumn{2}{c}{K2 PDCsap}  & 
\multicolumn{2}{c}{POLAR} \\ 
\cmidrule(lr){2-3} \cmidrule(lr){4-5}
BJD - 2400000 & Flux [e-/s] & Error [e-/s] & Flux  & Error \\ \hline
57994.328188 & 1654310.8 & 34.72 & 0.99997539  & 1.939\num{2e-5} \\
57994.348620 & 1654281.0 & 34.70 & 0.99999547  & 1.939\num{3e-5} \\
... & ... &     ... &   ... &   ... \\

\hline
\\\end{tabular}

\tablefoot{The complete table will be made available in machine-readable form in the online journal.
}

\end{table}

\section{Linear-scaled GLS periodograms} \label{Appendix:Linear_GLS}

\begin{figure}[!htb]
        \centering
        \includegraphics[width=\linewidth]{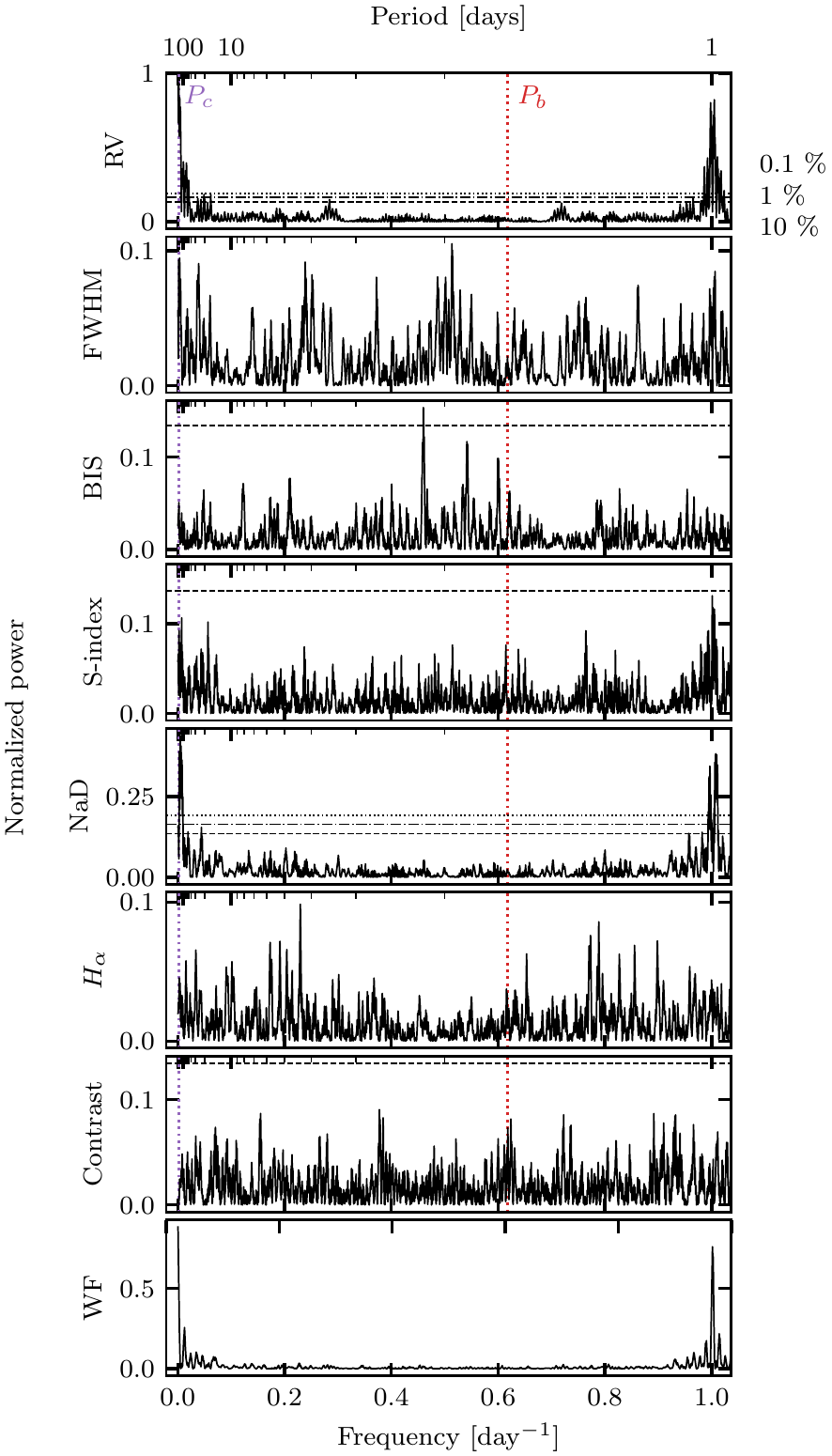}
        \caption{GLS periodograms of stellar activity indicators from HARPS RVs. The last panel represents the observational window function. 
}
    \label{fig:plot_fwhm_bis_si_HARPS_lin}
\end{figure}


\,

\begin{figure}[b!]
        \centering
        \includegraphics[width=\linewidth]{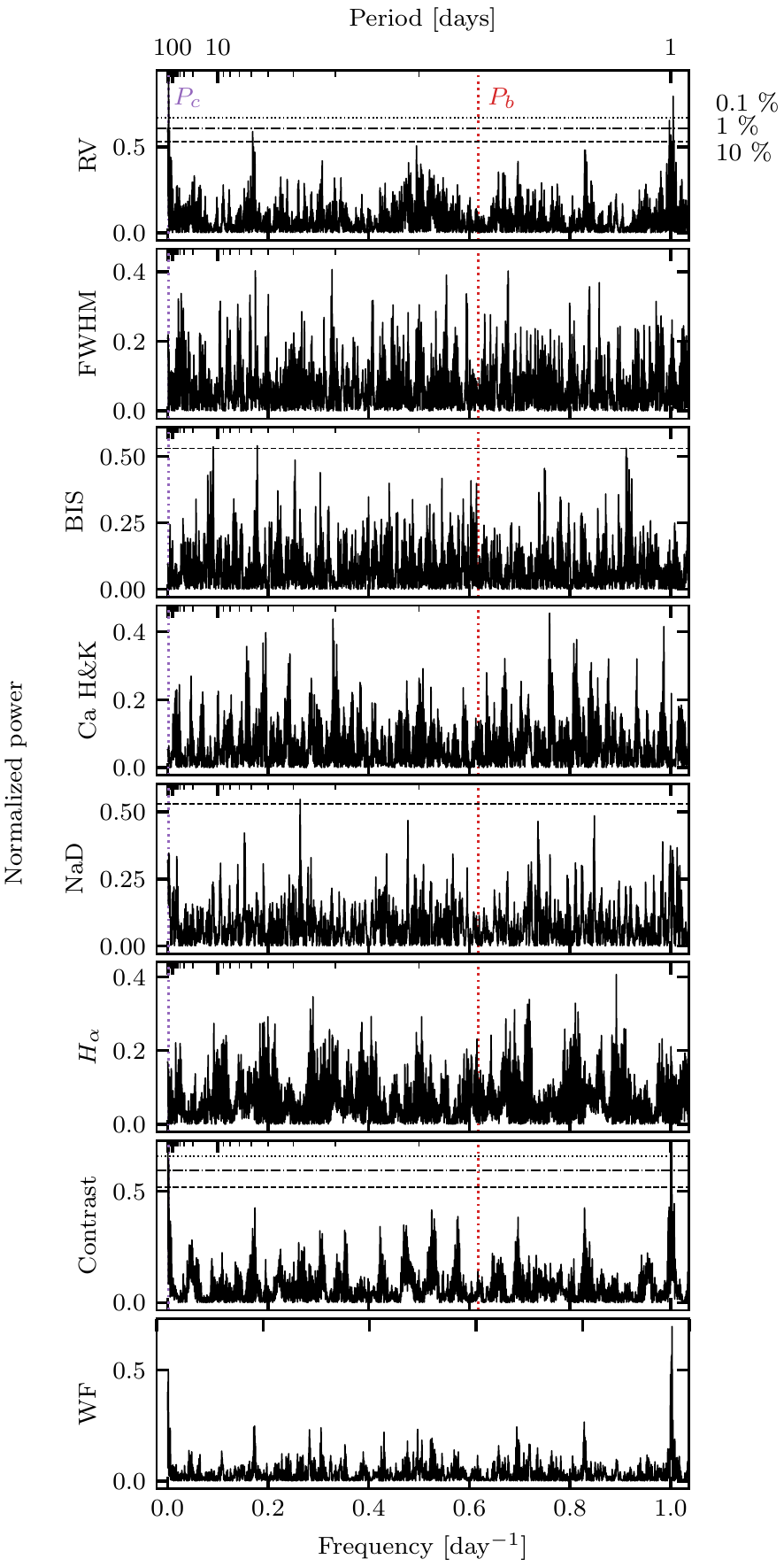}
        \caption{GLS periodograms of stellar activity indicators from CORALIE RVs. The last panel represents the observational window function.
}
    \label{fig:plot_fwhm_bis_si_CORALIE_lin}
\end{figure}

\clearpage

\section{Stellar parameter estimation} \label{Appendix:Ages}

As explained in Section \ref{section:stellar}, we chose to use PARAM to estimate the stellar mass, radius, and age of HD\,137496. In order to test the reliability of these estimates, we applied different tools to independently determine these parameters. In the following sections, we provide additional details on the analysis made with PARAM (Section \ref{Sub-appendix:PARAM}) and present the other approaches and their results (Sections \ref{Sub-appendix:SED fitting} - \ref{Sub-appendix:Chemical clocks}).

\subsection{Stellar modeling with PARAM} \label{Sub-appendix:PARAM}

Together with the parallax, $T_\textrm{eff}$, log $g,$ and [Fe/H] (Table \ref{table:stellar}), PARAM v1.5 can either perform a spectral energy distribution (SED) fit, using the magnitudes B, V, J, H, and K, or just consider the K\footnote{The K band is usually less affected by reddening, and the effect of low reddening can be included in the error bars.} magnitude and no SED fit. Both these methods provide very compatible age estimates, differing only by 0.1 Gyr, which is well within the error bars. However, our age estimate is very precise at 8.3 $\pm$ 0.7 Gyr. This is due to the proximity of the star to the turn-off region, in which the age has a greater impact on the shape of the isochrone, constraining the fit. The high precision on the derived metallicity additionally constrains the possible isochrones and results in such a small uncertainty. All the estimated values can be found in Table \ref{table:stellar}. 

\subsection{Stellar modeling with NextGen} \label{Sub-appendix:SED fitting}

\begin{figure}[!ht]
        \centering
        \includegraphics[width=\linewidth]{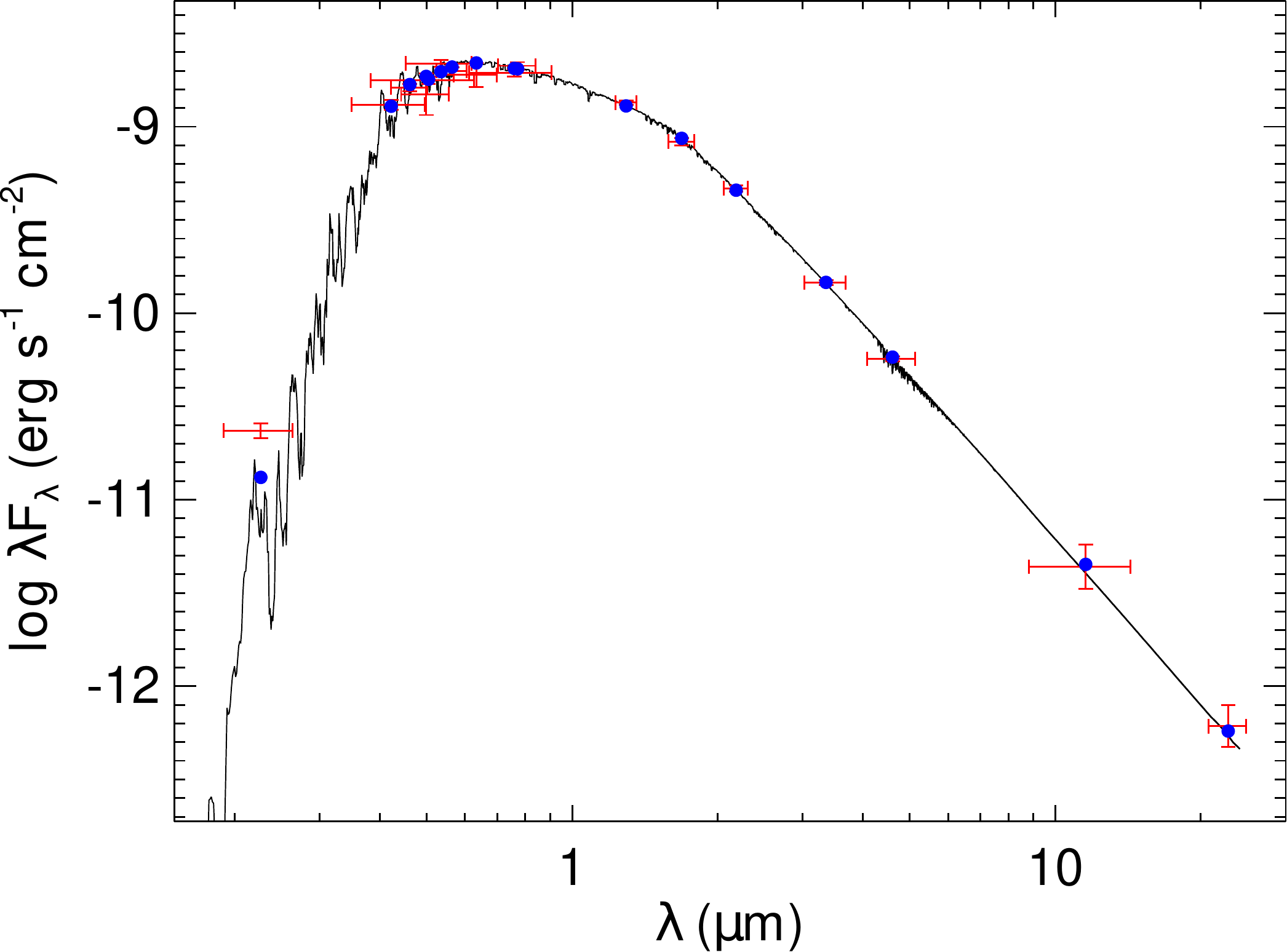}
        \caption{Spectral energy distribution of HD~137496. Red symbols represent the observed photometric measurements, where the horizontal bars represent the effective width of the passband. Blue symbols are the model fluxes from the best-fit NextGen atmosphere model (black).
}
    \label{fig:sed}
\end{figure}

We performed an analysis of the broadband spectral energy distribution (SED) of the star together with the {\it Gaia\/} EDR3 parallax \citep[with no systematic offset applied; see, e.g.,][]{StassunTorres:2021} following the procedures described in \citet{Stassun:2016} and \citet{Stassun:2017,Stassun:2018}. We pulled the $B_T V_T$ magnitudes from {\it Tycho-2}, the $BVgri$ magnitudes from {\it APASS}, the $JHK_S$ magnitudes from {\it 2MASS}, the W1--W4 magnitudes from {\it WISE}, the $G_{\rm BP} G_{\rm RP}$ magnitudes from {\it Gaia}, and the NUV magnitude from {\it GALEX}. Together, the available photometry spans the full stellar SED over the wavelength range 0.2--22~$\mu$m (see Figure~\ref{fig:sed}).  

We performed a fit using NextGen stellar atmosphere models, with the effective temperature ($T_{\rm eff}$), metallicity ([Fe/H]), and surface gravity ($\log g$) adopted from the spectroscopic analysis. The remaining free parameter is the extinction $A_V$, which we limited to the maximum line-of-sight value from the Galactic dust maps of \citet{Schlegel:1998}. The resulting fit (Figure~\ref{fig:sed}) has a best-fit $A_V = 0.29 \pm 0.04$ and a reduced $\chi^2$ of 1.3, but we note that the {\it GALEX\/} NUV flux indicates a mild level of chromospheric activity (see below). Integrating the (unreddened) model SED gives the bolometric flux at Earth, $F_{\rm bol} = 3.534 \pm 0.041 \times 10^{-9}$ erg~s$^{-1}$~cm$^{-2}$. Taking the $F_{\rm bol}$ and $T_{\rm eff}$ together with the {\it Gaia\/} parallax gives the stellar radius, $R_\star = 1.621 \pm 0.036$~R$_\odot$, which is consistent with a modestly evolved star. In addition, we can estimate the stellar mass from the empirical relations of \citet{Torres:2010}, giving $M_\star = 1.17 \pm 0.07$~M$_\odot$, which is consistent with the (less precise) estimate of $M_\star = 1.08 \pm 0.12$~M$_\odot$ obtained directly from $R_\star$ together with the spectroscopic $\log g$. 

Finally, we can use the star's NUV excess (Fig.~\ref{fig:sed}) to estimate a rotation period and age via empirical rotation-activity-age relations. The observed NUV excess implies a chromospheric activity of $\log R'_{rm HK} = -5.11 \pm 0.06$ via the empirical relations of \citet{Findeisen:2011}. This in turn implies a stellar rotation period of $P_{\rm rot} = 32 \pm 2$~d via the empirical relations of \citet{Mamajek:2008}, which could, in principle, be checked against a future $v\sin i$ and/or rotation period measurement, although these will be difficult given the very low level of activity of this evidently old star. Finally, the NUV estimated activity implies an age of $\tau_\star = 10.2 \pm 1.7$~Gyr via the empirical relations of \citet{Mamajek:2008}. This result is compatible with PARAM at a 1.1-$\sigma$ level, and the larger uncertainty can be expected since this estimate relies on an empirical relation.

\subsection{Stellar modeling with \texttt{PASTIS} } \label{Sub-appendix:PASTIS}

We also performed a Bayesian analysis of the stellar SED using the Planet Analysis and Small Transit Investigation Software \citep[\texttt{PASTIS};][]{Diaz2014, Santerne2015}.   \texttt{PASTIS} allows us to model the host star by fitting the SED to the BT-Settl atmosphere models \citep{Allard2012} and Dartmouth evolutionary tracks \citep{Dotter2008}.  For this, we used the $B$, $V$, $J$, $H,$ and $K$ magnitudes from Table \ref{table:stellar} as input.  The star was modeled for the effective temperature, T$_{\text{eff}}$; surface gravity, $\log$\,$g$; metallicity, [M/H]; distance, $d$; and color excess, \textit{E(B$-$V)}.  For these parameters, we used normal distributions defined by the spectroscopic values reported in Table \ref{table:stellar} as priors. Among other parameters (see Table~\ref{table:pastis}), we obtained a stellar age of 8.19 $\pm$ 0.69~Gyr, which is fully compatible with what was derived in the spectral analysis using PARAM. 

{\renewcommand{\arraystretch}{1.2}

\begin{table}[h]
\caption{\label{table:pastis} Stellar modeling with the PASTIS Bayesian code based on SED fitting.}
\centering
\begin{tabular}{lc}
\hline\hline
Parameter & Value \\

\hline

\hspace{3mm} Age (Gyr)   &   8.19 $\pm$ 0.69 \\
\hspace{3mm} $log\,g$ (cgs)    & 4.066 $\pm$ 0.020\\
\hspace{3mm} $R_{*} (R_{\odot}) $ & 1.580 $\pm$ 0.035 \\
\hspace{3mm} $M_{*} (M_{\odot})$  &  1.060 $\pm$ 0.023 \\
\hspace{3mm} $T_\text{eff}$  (K) & 5811 $\pm$ 58 \\
\hspace{3mm} E(B$-$V)  (mag) & 0.042 $\pm$ 0.026 \\
\hspace{3mm} [M/H] (dex) & -0.033 $\pm$ 0.040  \\
\hline

\end{tabular}
\end{table}}

\subsection{Stellar modeling with $q^2$} \label{Sub-appendix:q2}

We also estimated the mass, radius, and age of the host star through the use of the \texttt{Qoyllur-Quipu} (or $q^2$) Python package\footnote{Available at https://github.com/astroChasqui/q2} \citep{2014A&A...572A..48R} by inputting the T$_{\text{eff}}$, [Fe/H] and $\log$\,$g$ presented in Table \ref{table:stellar}. As suggested by \citep{2014A&A...572A..48R}, we first applied a -0.04 dex offset to the metallicity, and then we estimated $M_\star = 1.090 \pm 0.069$~$M_\odot$;
$R_\star = 1.617 \pm 0.235$~$R_\odot$ and age = 7.406 $\pm$ 1.328 Gyr by fitting the Yonsei–Yale isochrones \citep{2002ApJS..143..499K}. All these quantities are once again fully compatible with the PARAM analysis. The larger uncertainties can also be expected since, unlike the other approaches, this method does not take into account the parallax, SED, or any magnitudes.

\subsection{Chemical clocks} \label{Sub-appendix:Chemical clocks}

We applied the 3D formulas described in Table 10 of \citet{Delgado-19}, which also consider the variation in age produced by the effective temperature and iron abundance. The chemical clocks
[Y/Mg], [Y/Zn], [Y/Ti], [Sr/Ti], [Sr/Mg], [Y/Si], [Sr/Si], and [Y/Al] were used from which we obtain a weighted average of 6.7 Gyr (see Table \ref{table:Chem_1}). For the uncertainty, since the age estimates are derived from abundance ratios involving common species, assuming that all estimates are perfectly correlated we obtain a weighted error of 0.6 Gyr. However, the relations used to derive the age from each abundance ratio are derived empirically based on high-precision stellar abundances. As a consequence, the high precision of our abundance measurements of HD\,137496 results in a precise age estimate which might not reflect the true uncertainty of the method.

By using 1D formulas \citep[as described in Table 5 of][]{Delgado-19}, we obtain a similar age (see Table \ref{table:Chem_2}). The age from chemical clocks is lower than the age derived through isochrones with PARAM but still shows that this star is clearly older than the Sun. Interestingly, we find quite a high value of Li abundance (2.02\,dex) for an almost solar twin in terms of \teff\ and [Fe/H]. Several works have found that solar analogs hosting giant planets tend to present enhanced Li depletion \citep[e.g.,][and references therein]{Delgado-14,Figueira-14,Gonzalez-15} when compared to similar stars with no detected planets, but this is not the case of HD\,137496. However, this star has a low \logg\ value, indicating that it might be slightly evolved. The works by \citet{Delgado-14} and \citet{Baumann-10} show that stars with \logg  $\lesssim$ 4.2\,dex increase their Li as \logg\ decreases, regardless of the presence of planets. The higher Li abundances of these somewhat evolved stars could be explained by the fact that they were hotter on the main sequence and did not deplete as much Li due to their thinner convective envelopes. The Li abundance, \logg\ value and the radius larger than the Sun support the fact that this star is entering the subgiant phase. This indicates an older age than the 6.8\,$\pm$\,0.7 Gyr derived from the chemical clocks and is more compatible with the 8.3\,$\pm$\,0.7 Gyr derived using PARAM. We also note that our chemical clock analysis assumes that the star is in on the main sequence. \citet{liu2019} showed that atomic diffusion can significantly alter the measured abundances when a star leaves the main sequence and thus might bias the age estimate provided by chemical clocks. We thus favored the PARAM age estimate over the one provided by the chemical clocks.

{\renewcommand{\arraystretch}{1.2}

\begin{table}[h]
\caption{\label{table:Chem_1}Age estimation from 3D relations \citep[more information can be found in ][]{Delgado-19}.}
\centering
\begin{tabular}{cc}
\hline\hline
Formula&Age\\

\hline

\hspace{3mm} $T_\text{eff}$ + [Fe/H] + \text{[Sr/Si]} &    7.41 $\pm$ \;        2.31 \\
\hspace{3mm} $T_\text{eff}$ + [Fe/H] + \text{[Sr/Si]} &    7.41 $\pm$ \;        2.16 \\
\hspace{3mm} $T_\text{eff}$ + [Fe/H] + \text{[Sr/Mg]} &    7.67 $\pm$ \;        1.72 \\
\hspace{3mm} $T_\text{eff}$ + [Fe/H] + \text{[Y/Mg]} &    6.72 $\pm$ \; 1.21 \\
\hspace{3mm} $T_\text{eff}$ + [Fe/H] + \text{[Y/Si]} &    6.10 $\pm$ \; 1.67 \\
\hspace{3mm} $T_\text{eff}$ + [Fe/H] + \text{[Y/Ti]} &    6.34 $\pm$ \; 1.46 \\
\hspace{3mm} $T_\text{eff}$ + [Fe/H] + \text{[Y/Zn]} &    6.66 $\pm$ \; 1.38 \\

\hline

\end{tabular}

\end{table}}

{\renewcommand{\arraystretch}{1.2}

\begin{table}[h]
\caption{\label{table:Chem_2}Age estimation from 1D relations \citep[more information can be found in ][]{Delgado-19}.}
\centering
\begin{tabular}{cc}
\hline\hline
Formula&Age\\

\hline

\hspace{3mm} \text{[Sr/Ti]} &    6.88 $\pm$ \;  2.31 \\
\hspace{3mm} \text{[Sr/Mg]} &    6.82 $\pm$ \;  2.06 \\
\hspace{3mm} \text{[Sr/Si]} &    6.58 $\pm$ \;  1.21 \\
\hspace{3mm} \text{[Sr/Zn]} &    6.34 $\pm$ \;  1.09 \\
\hspace{3mm} \text{[Y/Al]} &    6.61 $\pm$ \;   0.50 \\
\hspace{3mm} \text{[Y/Mg]} &    6.86 $\pm$ \;   0.76 \\
\hspace{3mm} \text{[Y/Si]} &    6.97 $\pm$ \;   0.91 \\
\hspace{3mm} \text{[Y/Ti]} &    6.91 $\pm$ \;   0.94 \\
\hspace{3mm} \text{[Y/Zn]} &    6.58 $\pm$ \;   0.82 \\

\hline

\end{tabular}

\end{table}}

\section{Selection of priors} \label{Appendix:MCMC_priors}

\subsection{Instrumental parameters}

For the systemic velocity $\gamma_{RV,HARPS,}$ we considered a Gaussian prior with the median and standard deviation matching that of the HARPS data. Regarding the jitter parameters, we chose Jeffreys priors evaluated between $10^{-2}$ and 10 times the median error bar for each of the datasets. The prior of the RV shift between the CORALIE and HARPS data was set as a uniform prior between the median plus and minus the square root of the sum of the variances computed from all data after 2458673 BJD, when both instruments observed the target.

\subsection{Stellar parameters}

For the limb-darkening coefficients, we use the PyLDTk\footnote{Available at \url{https://github.com/hpparvi/ldtk}} \citep{Parviainen2015} toolkit. By applying this method for the \textit{Kepler} bandpass and taking the previously determined $T_\text{eff}$ and log $g$ as inputs, the result for the limb-darkening model coefficients of the quadratic model are $u1 = 0.5324 \pm 0.0061$ and $u2 = 0.1056 \pm 0.0091$. We used these estimates as priors. Finally, for the stellar density, we used the estimate obtained by the stellar characterization analysis (see Section \ref{section:stellar}).

\subsection{Planetary parameters }

With respect to the planetary periods, we have broad Jeffreys priors set between the shortest orbital period for a known planet \cite[0.09 days - ][]{2011Sci...333.1717B} and the time span of our data (1400 days). With regards to the RV semi-amplitudes, $K$, we again set broad priors. For planet b, we considered a uniform prior between 0 and 50 m/s, so as to safely cover all realistic scenarios for a planet with the observed transit depth. For planet c, we considered a uniform prior between 0 and 444 m/s, the range of RVs in our HARPS dataset. 

Joint priors are used in this work as a way to better convey our prior knowledge of the system. They consist of the conversion of one set of jumping parameters into a different but equivalent set of parameters, allowing for a more physically motivated prior while helping the convergence of the MCMC analysis.

One example is the joint prior on the jumping parameter $e \cos \omega $ and $e \sin \omega$. It is not possible to set two independent priors on these two quantities that would be equivalent to a prior on $e$ between 0 and 1 and a prior on $\omega$ between -$\pi$ and $\pi$. To solve this problem, we used a joint prior on  $e \cos \omega $ and $e \sin \omega$ which converts this set of parameters into $e$ and $\omega$. We can then set priors directly on $e$ and $\omega$, while $e \cos \omega $ and $e \sin \omega$ remain the jumping parameters, which is important for a better convergence \cite{eastman2013exofast}. For planet c, we used a uniform prior between $-\pi$ and $\pi$ for $\omega$ and a Beta distribution (with $a = 0.867$ and $b = 3.03$) for $e$ \citep{kipping2013b}. In the case of planet b, due to its short orbital period and the age of the system, the orbit is expected to be circular. Since from Equation (1) of \cite{dobbs-dixon2004} we can expect a timescale associated with an eccentricity damping of 1.44 Myr, for this calculation we assume a modified tidal quality factor $Q'_p$ of 100.

Another joint parameter used in this work is the joint parameter on $P$ and $t_{ic}$ for planet c. We do not have prior knowledge of any of these two parameters; to reflect this lack of constraints, we need to set broad uninformative priors on both these quantities. However, by definition $t_{ic}$ is a degenerate variable as the values of $t_{ic}$ and $t_{ic}$ + $P$ are equivalent in our model. To avoid this degeneracy, we used a joint prior that transforms the set of parameters $P$ and $t_{ic}$, into $P$ and $\phi$, where $\phi$ is the orbital phase defined by $\phi$ = ($t_{ic}$ - $t_{ref}$) / $P$. In this equation, $t_{ref}$ is a reference time arbitrarily chosen as the rounded value of the first RV measurement. With this new set of parameters, we can easily set broad and uninformative priors while removing the degeneracy. We chose a Jeffreys prior between 0.09 and 1400 days on $P$ and a uniform prior between 0 and 1 on $\phi$. 

Our final joint prior is designed to translate the prior knowledge that a planet is transiting. From our previous analysis on the photometric (Section \ref{section:photometry}) data, we know that planet b is transiting. The S/R of the transit of planet b is low (see Fig. \ref{fig:plot_3subs_phase_curves}), and it is too easy for the MCMC exploration to wander off into scenarios where planet b is not transiting. We can significantly improve the MCMC convergence by limiting the parameter space to a region where planet b is transiting. The parameters used in this model are the period, $P$, impact parameter, $b$, radius ratio, $R_p/R_*$, and stellar density,  $\rho_*$ \citep{seager2003unique}. Both $\rho_*$ or $a/R_*$, together with $P$, $b$ and $R_p/R_*$ would accurately model the transit. In this work, since we have strong constraints on $\rho_*$ from the stellar analysis (Section \ref{section:stellar}), we can use this information to achieve a better convergence on the MCMC and improve the results on our posterior distribution of $b$. If not for our prior information on the stellar density, a degeneracy could arise between $\rho_*$  and $b$, as both are capable of affecting the transit duration. By using a joint prior, we compute $b$ alongside $R_p/R_*$, allowing us to always determine if the planet is transiting for each set of parameters. By imposing the conditions 0 < $R_p/R_*$ < 1 and 0 < $b$ < 2, we include every non-grazing and grazing scenario.


\section{Early run: Prior and posterior}\label{Appendix:Early_run}

For the joint fit used to derive the final estimates of the parameters of the system (Section \ref{sec:joint}), we chose broad prior distributions (see Table \ref{table:results_mcmc} and Appendix \ref{Appendix:MCMC_priors}) that reflected any prior knowledge on the system and did not unjustifiably constrain the parameters. However, in the attempts that we made, when we drew the initial position of the MCMC chains from these priors, we did not converge to the best solution. From the preliminary analysis of the light curve (see Section \ref{section:photometry}) and the RV (see Section \ref{section:rvs}), we obtained a rough knowledge of the best values for $P_b$, $t_{ic,b}$, $P_,c$ and $K_c$. This knowledge could not be used to define the prior distributions since it arose from the data that we were analyzing. However, to circumvent the problem, we carried out a two-stage analysis: a first run to converge to the best region of the parameter space and a second to derive robust estimates for the value of the parameters. So, we did a first run, which used the knowledge of $P_b$, $t_{ic, b}$, $P_c,$ and $K_c$ extracted from our preliminary analyses to define their prior distribution (see Table \ref{table:app_early_run}). We then performed a second run (Section \ref{sec:joint}) that used the broader priors defined in Table \ref{table:results_mcmc} but the posterior distributions of the parameters listed in Table \ref{table:app_early_run}, coming from the first run, to draw the initial values of the MCMC chains.

\newpage

{\renewcommand{\arraystretch}{1.5}

\begin{table}[h!]
\caption{\label{table:app_early_run}
Prior and posterior distributions for our first run designed to converge to the best region of the parameter space quickly ($\mathcal{U}$ - Uniform; $\mathcal{G}$ - Gaussian; $\mathcal{J}$ - Jeffreys; $\mathcal{B}$ -  beta distribution).}
\centering 
\begin{tabular}{lcc}
\hline\hline
Parameter & Prior & Posterior\tablefoottext{a}\!\\

\hline

\hspace{3mm} Stellar density $\rho_*$ [$\textrm{g.cm}^{-3}$] & $\mathcal{G}$ [0.2590, 0.0150] &  $\mathcal{G}$ [0.257, 0.02] \\

\hline

\multicolumn{3}{c}{\textbf{HD\,137496\,b}}  \\
\hline

\hspace{3mm} Period $P_b$ [days] & $\mathcal{G}$ [1.621, 0.0005]  & $\mathcal{G}$ [1.6211, 0.0001]  \\
\hspace{3mm} RV semi-amplitude $K_b$ [\ms] & $\mathcal{U}$ [0.0, 50.0] & $\mathcal{G}$ [2.14, 0.3]  \\
\hspace{3mm} Radius ratio $R_{p, b}/R_*$  & $\mathcal{U}$ [0.0, 1.0] & $\mathcal{G}$ [0.0076, 0.0003]  \\
\hspace{3mm} Time of inferior conjunction $t_{ic, b}$ [\text{BJD{\text -}2450000}] & $\phi$ : $\mathcal{U}(0.08, 0.006)$\tablefoottext{b} & $\mathcal{G}$ [58039.132, 0.003] \\
\hspace{3mm} Cos inclination $\cos i_b$ [º] & - & $\mathcal{G}$ [0.072, 0.06] \\

\hline

\multicolumn{3}{c}{\textbf{HD\,137496\,c}}  \\
\hline

\hspace{3mm} Period $P_c$ [days] & $\mathcal{J}$ [0.1, 740] & $\mathcal{G}$ [481.6, 3.0] \\
\hspace{3mm} RV semi-amplitude $K_c$ [\ms] & $\mathcal{G}$ [230, 100] & $\mathcal{G}$ [221.26, 0.9] \\
\hspace{3mm} Time of inferior conjunction $t_{ic, c}$ [\text{BJD{\text -}2450000}] & $\phi$ : $\mathcal{U}(0.0, 1.0)$\tablefoottext{b} &  $\mathcal{G}$ [58629.22, 0.5] \\

\hline

\end{tabular}

\tablefoot{
\tablefoottext{a}{Starting distribution from which the samples were drawn for the first step in our MCMC analysis.} \\
\tablefoottext{b}{As defined in Table \ref{table:results_mcmc}.} \\

}

\end{table}}

\end{document}